\documentclass{article}

\usepackage{arxiv}

\usepackage[utf8]{inputenc} 
\usepackage[T1]{fontenc}    
\usepackage{hyperref}       
\usepackage{url}            
\usepackage{booktabs}       
\usepackage{amsfonts}       
\usepackage{nicefrac}       
\usepackage{microtype}      
\usepackage{lipsum}

\usepackage{listings}
\usepackage{rotating}
\usepackage{multirow}
\usepackage{xspace}
\usepackage{ascmac}

\usepackage{color}

\newif\ifdraft
\draftfalse 

\newif\ifcheck
\checkfalse 

\newif\ifwithfigure
\withfiguretrue 

\newcommand{\secref}[1]{Section~\ref{#1}}

\newcommand{\figref}[1]{Fig~\ref{#1}}
\newcommand{\tabref}[1]{Table~\ref{#1}}
\newcommand{\eqnref}[1]{Eq.~\ref{#1}}

\newcommand{\hw}[2]{\parbox{#1em}{\centering #2}}

\newcommand{\figcontent}[5]{
  \begin{center}
    \includegraphics[#1]{#2}
    \caption{#3}
    \label{#4}
    #5
  \end{center}
}

\newcommand{\figframe}[2][t]{
  \begin{figure}[#1]
        #2
  \end{figure}
}

\newcommand{\fig}[6][t]{
  \figframe[#1]{\figcontent{#2}{#3}{#4}{#5}{#6}}
}

\newcommand{\eqn}[1]{\begin{eqnarray}#1\end{eqnarray}}

\newcommand{\elist}[1]{
  \begin{enumerate}
    #1
  \end{enumerate}
}
\newcommand{\ilist}[1]{
  \begin{itemize}
    #1
  \end{itemize}
}
\newcommand{\dlist}[1]{
  \begin{description}
    #1
  \end{description}
}
\newcommand{\dlbr}[1]{\mbox{}\\[#1]}

\newcommand{\dltxt}{\bf}

\newcommand{\tablcontent}[4]{
  \renewcommand\thefootnote{*\alph{footnote}}
  \begin{center}
      \begin{center}
      \caption{\textbf{#1}}
      \label{#2}
      \begin{tabular}{#3}
        #4
    \end{tabular}
    \end{center}
  \end{center}
}

\newcommand{\tabl}[7][t]{
  \begin{table}[#1]
      \tablcontent{#2}{#3}{#4}{#5}
      \begin{flushleft}#6\end{flushleft}
      \vspace{#7}
  \end{table}
}

\newcommand{\cnum}[1]{\textcircled{\fontsize{5.7pt}{0pt}\selectfont#1}}

\newcommand{\boldTitle}[1]{\noindent{\textit{\textbf{#1:\ }}}}

\newcommand{\ParmoSense}{ParmoSense\xspace}
\newcommand{\ParmoSenseDashboard}{ParmoSense Dashboard\xspace}
\newcommand{\ParmoSenseClient}{ParmoSense Client\xspace}
\newcommand{\ParmoSenseServer}{ParmoSense Server\xspace}

\newcommand{\ScenarioManager}{Scenario manager\xspace}
\newcommand{\DataManager}{Data manager\xspace}
\newcommand{\InternalSystem}{Scenario instance\xspace}
\newcommand{\InternalSystems}{Scenario instances\xspace}

\newcommand{\ScenarioTools}{Scenario tools\xspace}
\newcommand{\ScenarioEditor}{Scenario editor\xspace}

\newcommand{\DataTools}{Data tools\xspace}

\newcommand{\organizer}{organizer\xspace}
\newcommand{\organizers}{organizers\xspace}
\newcommand{\participant}{participant\xspace}
\newcommand{\participants}{participants\xspace}

\newcommand{\SensingFuncs}{sensing functions\xspace}
\newcommand{\MotivatingFuncs}{motivating functions\xspace}
\newcommand{\ProcessingFuncs}{processing functions\xspace}

\newcommand{\DistributingPhase}{Distributing-phase\xspace}
\newcommand{\SensingPhase}{Sensing-phase\xspace}
\newcommand{\ProcessingPhase}{Processing-phase\xspace}

\newcommand{\PSfunctions}{functions\xspace}

\newcommand{\PSFunctions}{Functions\xspace}

\title{ParmoSense:\\A Scenario-based Participatory Mobile Urban Sensing Platform\\with User Motivation Engine}

\author{
  Yuki Matsuda\textsuperscript{1,2,3,$\dagger$,}\thanks{\url{http://yukimat.jp/}}\ ,
  Shogo Kawanaka\textsuperscript{1,2,4},
  Hirohiko Suwa\textsuperscript{1,2},
  Yutaka Arakawa\textsuperscript{5},
  Keiichi Yasumoto\textsuperscript{1,2}\\
    \textsuperscript{1} Graduate School of Science and Technology, Nara Institute of Science and Technology, Ikoma, Nara, Japan\\
    \textsuperscript{2} Center for Advanced Intelligence Project AIP, RIKEN, Chuo, Tokyo, Japan\\
    \textsuperscript{3} PRESTO, JST, Chiyoda, Tokyo, Japan\\
    \textsuperscript{4} Research Fellowship for Young Scientists, JSPS, Chiyoda, Tokyo, Japan\\
    \textsuperscript{5} Graduate School and Faculty of Information Science and Electrical Engineering,\\
    Kyushu University, Fukuoka, Fukuoka, Japan\\
  \textsuperscript{$\dagger$} \texttt{yukimat@is.naist.jp}
}

\begin{document}
\maketitle

\begin{abstract}
Rapid proliferation of mobile devices with various sensors have enabled \emph{Participatory Mobile Sensing (PMS)}. Several PMS platforms provide multiple functions for various sensing purposes, but they are suffering from the open issues: limited use of their functions for a specific scenario/case and requiring technical knowledge for organizers.
In this paper, we propose a novel PMS platform named \emph{ParmoSense} for easily and flexibly collecting urban environmental information.
To reduce the burden on both organizers and participants, in ParmoSense, we employ two novel features: \textit{modularization of functions} and \textit{scenario-based PMS system description.}
For modularization, we provide the essential PMS functions as modules which can be easily chosen and combined for sensing in different scenarios. The scenario-based description feature allows organizers to easily and quickly set up a new participatory sensing instance and participants to easily install the corresponding scenario and participate in the sensing.
Moreover, ParmoSense provides GUI tools as well for creating and distributing PMS system easily, editing and visualizing collected data quickly.
It also provides multiple functions for encouraging participants' motivation for sustainable operation of the system.
Through performance comparison with existing PMS platforms, we confirmed ParmoSense shows the best cost-performance in the perspective of the workload for preparing PMS system and varieties of functions.
In addition, to evaluate the availability and usability of ParmoSense, we conducted 19 case studies, which have different locations, scales, and purposes, over 4 years with cooperation from ordinary citizens. Through the case studies and the questionnaire survey for participants and organizers, we confirmed that ParmoSense can be easily operated and participated by ordinary citizens including non-technical persons.
\end{abstract}


\section{Introduction}
\label{sec:introduction}
Mobile devices come equipped with various sensors including GPS, inertial sensors, environment sensors, camera, microphone and so on.
The rapid widespread of such mobile devices 
have enabled Participatory Mobile Sensing (PMS)~\cite{bib:burke_participatory_2006, bib:campbell_acm_2006, bib:paulos_citizenScience_2008}. PMS systems are based on crowdsourcing technology, where data in a wide geographical area can be collected efficiently and at low cost by leveraging sensors on mobile devices carried by ordinary citizens. Many applications can utilize collected urban data to bring various benefits to our daily lives. For instance, because PMS systems use common
devices, they can be easily used to collect data for urban analysis~\cite{bib:morishita_ubicomp_2015}, office management~\cite{bib:konis_occupantMobileGateway_elsevier_2017}, healthcare~\cite{bib:rolt_collega_iscc_2016}, and education~\cite{bib:heggen_ubicomp_2012,bib:kok_journalofPerCom_2014}. In addition, since PMS systems can be applied to any region in which people stay or pass, they are very effective for collecting geospatial data over a wide area. In urban environment for example, data such as illuminance of the road at night~\cite{bib:matsuda_ubicomp_2014}, noise levels in the city~\cite{bib:kanjo_NoiseSPY_2010, bib:maisonneuve_noise_2010}, and air pollution degrees~\cite{bib:mendez_pSense_2011, bib:zheng_uAir_2013} can be collected.

PMS is a sensing mechanism based on the voluntarism of general people. In other words, the sustainability of the system is a critical challenge in real-world operations. As an idea to enhance this sustainability, the mutual linkage with the local community where ecosystems are already formed can be considered. For example, the civic cooperation that people work with government, universities, companies, etc. to promote community development spreads globally. Especially in recent years, CivicTech which combines ICT and civic cooperation is gathering attention, e.g., FixMyStreet~\cite{bib:fixmystreet_org}. In our study, we focus on PMS systems which can be used in CivicTech communities.

To realize PMS systems in the real world for broad urban environment analysis, we believe that a platform that can be easily and quickly customized by organizers to perform a variety of sensing tasks, and that is easy to set-up and run on the participants' smartphones, is essential.
However, when we investigated the functions implemented in existing PMS platforms~\cite{bib:ferreira_aware_frontiers_2015, bib:haoyi_sensus_ubicomp_2016, bib:ra_medusa_mobisys_2012, bib:funf_social_fMRI_Percom_2011, bib:sakamura_minaqn_ubicomp_2015, bib:mishima_kokopin_maed_2013, bib:tangmunarunkit_ohmage_acmTran_2015, bib:brunette_opendatakit_hotmobile_2013, bib:wang_GPSelector_www}, we found two main challenges in using these platforms for broad urban environment analysis: \emph{C1: Limited support of essential functions} and \emph{C2: Difficulty of system construction and operation}.

Regarding \emph{C1}, existing platforms tend to focus on specific sensing purposes, e.g., urban transport data sensing, and therefore support limited functions. Because the purpose of sensing differs among organizers of urban sensing, the necessary functions,  i.e., sensing function, incentive mechanism, task request control, and data processing method, will also differ depending on the purposes. Thus, in the ideal PMS platform, flexibility to adapt the platform to perform sensing for various purposes is mandatory.
Moreover, motivating participants is an important aspect of PMS since participatory sensing relies on voluntary participation of ordinary citizens~\cite{bib:arakawa_gamification_2016}, but we found that these platforms do not implement it enough.
And regarding \emph{C2},
The platforms require a high level of technical skill for users. For example, some platforms require programming skills for organizers and data processing skills for participants. In order to open the door of participatory sensing to non-technical users, it is necessary to ensure that PMS systems can be easily constructed and operated by both organizers and participants.

In this study, we designed and built a novel PMS platform called \emph{ParmoSense}, for easily and flexibly collecting urban environmental information for various purposes by overcoming the challenges mentioned above. To achieve this, we employ two features: \textit{modularization of functions} and \textit{scenario-based PMS system description.} We provide various functions essential for PMS systems such as sensing functions, motivating functions for participants, and processing functions for collected data, and allow organizers to combine these modularized functions freely through a GUI web application. We call a combination of these modularized functions a \textit{scenario}. Once a scenario has been created, participants can download it onto the \ParmoSenseClient application and run it without doing any further setup or processing tasks. Thus, participants can contribute to many different sensing tasks without installing multiple-applications or performing complicated tasks which require technical skills.

To evaluate the superiority of ParmoSense, we compared a performance with existing PMS platforms. First, we confirmed that ParmoSense provides a higher variety of functions than the existing PMS platforms, which solves challenge \emph{C1}. We also found that ParmoSense eases organizers to prepare PMS systems. It belongs to the lowest preparation workload group among those platforms, which solves challenge \emph{C2}. From both perspectives, therefore, ParmoSense shows the best cost-performance.
In addition,
to evaluate the availability and usability of ParmoSense in the real-world,
we conducted the 19 practical case studies with ordinary citizens including non-technical people.
We confirmed that ParmoSense can deal with various sensing targets, organizers, and participants, in real environments.

Our contributions in this paper are as follows:
\elist{
  \item We designed a PMS platform, named ParmoSense, which allows general people to easily operate PMS systems with scenario-based system construction regardless technical skills.
  \item We organized functionality requirements through the survey of existing PMS platforms, and implemented all functions as combinable modules.
  \item We confirmed that ParmoSense shows the best cost-performance in the perspective of varieties of functions and preparation workload through the evaluation on comparison with nine existing platforms.
  \item We confirmed the availability and usability of ParmoSense through the 19 practical case studies in the real world, and interviews with participants and organizers of sensing tasks.
}

The rest of this paper is organized as follows: 
In \secref{sec:related-work}, we survey the existing PMS platforms and systems, and organize required functions for PMS systems and skill requirement for participants and organizers.
In \secref{sec:architecture}, we describe the concept and the architecture of ParmoSense and provided functions on ParmoSense.
To evaluate ParmoSense, we compare the functionality and performance with existing PMS platforms in \secref{sec:evaluation}, and conduct the 19 practical case studies with general people including non-technical persons in \secref{sec:case_studies}.
\secref{sec:conclusion} concludes this paper, and we will discuss limitation and future challenges of ParmoSense.


\section{Related work and challenges}
\label{sec:related-work}

This section is devoted to clarifying what kind of functions are necessary for PMS systems and 
what kind of skills are required for users (organizers and participants) of PMS systems.
We first organized the necessary functions into three categories: sensing functions, motivating functions and processing functions.
Then, we investigated the functions implemented in existing PMS platforms~\cite{bib:ferreira_aware_frontiers_2015, bib:haoyi_sensus_ubicomp_2016, bib:ra_medusa_mobisys_2012, bib:funf_social_fMRI_Percom_2011, bib:sakamura_minaqn_ubicomp_2015, bib:mishima_kokopin_maed_2013, bib:tangmunarunkit_ohmage_acmTran_2015, bib:brunette_opendatakit_hotmobile_2013}. The results are summarized in \tabref{tab:funcs-overview}.
The skills required by organizers and participants in existing PMS platforms are shown in \tabref{tab:skill-requirement-overview}.

\renewcommand{\arraystretch}{1.3}
\newcommand{\scTriangle}[1]{\hspace{0.6em}\small $\triangle$\footnotemark[#1]}

\tabl[t]{Overview of supported functions.}{tab:funcs-overview}{c|cc|ccc|ccc}{\hline\hline
    \multirow{4}{*}{Platforms}    & \multicolumn{8}{c}{\PSFunctions}       \\ \cline{2-9}
          & \multicolumn{2}{c|}{Sensing functions }  
                & \multicolumn{3}{c|}{Motivating functions } 
                & \multicolumn{3}{c}{Processing functions }      \\ \cline{2-9}
            & \hw{3.5}{F1}  & \hw{3.5}{F2}  & \hw{2.5}{F3}  & \hw{2.5}{F4} 
             & \hw{2.5}{F5}  & \hw{2.5}{F6}  & \hw{2.5}{F7}  & \hw{2.5}{F8}     \\[-2.3mm]
                & {\scriptsize Implicit-sensing}  
                & {\scriptsize Explicit-sensing}
                & {\scriptsize Request} 
                & {\scriptsize Reward}    
                & {\scriptsize Feedback} 
                & {\scriptsize Editing}  
                & {\scriptsize Browsing} 
                & {\scriptsize Export} \\ \hline
    AWARE~\cite{bib:ferreira_aware_frontiers_2015}
          & \checkmark     & \scTriangle{2}
          & \scTriangle{5} &               & \scTriangle{7}
          &                & \checkmark    & \checkmark  \\
    Sensus~\cite{bib:haoyi_sensus_ubicomp_2016}
          & \checkmark   & \scTriangle{3}
          &              &                &         
          &              & \checkmark     & \scTriangle{8} \\
    Medusa~\cite{bib:ra_medusa_mobisys_2012}
          & \checkmark      & \scTriangle{4}
          & \scTriangle{5}  & \scTriangle{6}  & 
          &                 & \checkmark   &       \\\hline
    Funf~\cite{bib:funf_social_fMRI_Percom_2011}
          & \checkmark    & \scTriangle{4}
          &               &               & 
          & \checkmark    & \checkmark    & \checkmark  \\
    MinaQn~\cite{bib:sakamura_minaqn_ubicomp_2015}
          & \scTriangle{1}  & \scTriangle{2}
          & \scTriangle{5}  &         & \checkmark
          &         & \checkmark    &         \\
    KOKOPIN app~\cite{bib:mishima_kokopin_maed_2013}
          & \scTriangle{1}  & \scTriangle{3}
          &                 &                 & \checkmark
          &                 & \checkmark      & \checkmark    \\ \hline
    Ohmage~\cite{bib:tangmunarunkit_ohmage_acmTran_2015}
          & \scTriangle{3}  & \checkmark
          &                 &         & \scTriangle{7}
          & \checkmark      & \checkmark    & \checkmark  \\
  OpenDataKit~\cite{bib:brunette_opendatakit_hotmobile_2013}
          & \scTriangle{1}  & \checkmark
          &                 &               & 
          & \checkmark      & \checkmark    & \checkmark  \\ 
  GP-Selector~\cite{bib:wang_GPSelector_www}
          & \checkmark    & \checkmark
          & \checkmark    & \scTriangle{6} & 
          &               &                &            \\ \hline
  ParmoSense
          & \checkmark    & \checkmark
          & \checkmark    & \checkmark    & \checkmark
          & \checkmark    & \checkmark    & \checkmark  \\\hline
}{
\scriptsize
\footnotemark[1] Location (GPS) only supported.\hspace{0.5em}
\footnotemark[2] Media upload (photo, sound, etc.) limited.\hspace{0.5em}
\footnotemark[3] Raw data collection unsupported.\hspace{0.5em}
\footnotemark[4] Questionnaire unsupported.\\[0.5mm]
\footnotemark[5] Static request only supported.\hspace{0.5em}
\footnotemark[6] Monetary incentives only supported.\hspace{0.5em}
\footnotemark[7] Feedback of data collected by oneself only supported.\hspace{0.5em}
\footnotemark[8] command-line tool is required.
}{0mm}

\subsection{Functions essential in PMS platform}

\subsubsection{Sensing functions}

Sensing is an essential part of PMS systems. We define sensing functions as those that allow the organizer to specify what kind of sensors to use, and how to collect sensing data from the urban environment.
There are two different ways of sensing: implicit (F1) and explicit (F2) sensing.

\begin{description}
  \item[\dltxt Implicit-sensing (F1):]
    Uses sensors embedded in mobile devices. It is mainly used for collecting urban environmental data without actively involving the participant, i.e., implicit sensing.
  \item[\dltxt Explicit-sensing (F2):]
    Used for collecting data generated by human behavior, e.g., photos, voice, and questionnaires. It is used for collecting urban environmental data through directly involving participants, i.e., explicitly, and locally.
\end{description}

AWARE~\cite{bib:ferreira_aware_frontiers_2015} provides a platform for both implicit and explicit sensing. For implicit-sensing, the organizer can choose which smartphone sensors to use from a web UI. They can also configure the detailed settings such as sensing interval. For explicit-sensing, AWARE allows the organizer to distribute questionnaires manually. 

Ohmage~\cite{bib:tangmunarunkit_ohmage_acmTran_2015} supports explicit-sensing, by allowing participants to post reports by themselves. Several report formats are accepted  such as single/multiple selections, free text, multimedia (e.g., photo, sound) and so on.
Ohmage also supplements collected data through implicit-sensing. Specifically, it can be used to record the transportation status (e.g., still, walk, run) of a participant.

The other conventional platforms however, tend to focus more on either implicit- or explicit-sensing (as shown in \tabref{tab:funcs-overview}), and provide limited functionality for the other form of sensing. As mentioned before, the two sensing methods have many differences such as data type that can be collected, and the data's features. Thus, with these platforms, it is difficult to supplement the collected data due to severe restrictions in sensing functions.

In this paper, in order to realize a flexible sensing platform, we aim to provide functions  for both sensing methods comprehensively and support the organizers' ability to easily choose and combine functions.\\[-2mm]

\subsubsection{Motivating functions}

Since PMS relies on voluntary participation of ordinary citizens, it is essential to not only focus on acquiring users, but also on motivating them to continue participating in the sensing tasks, i.e., to support user retention and activation~\cite{bib:gamification_co_fatigue_2012}.
Motivating functions allow the organizer to define the methods for motivating participants.
Some of the conventional platforms use outside stimuli to support participant motivation and engagement.
We classify these outside stimuli as follows:

\begin{description}
  \item[\dltxt Request (F3):]
    This is urging behavior by explicitly requesting participation.
    There are many methods of sending requests.
    The most common method is the static/dynamic request, which includes providing a task list, issuing notifications and so on.
    Other methods such as Audition~\cite{bib:ra_medusa_mobisys_2012}, 
    and Reverse-auction~\cite{bib:Francesco_incentive_2015, bib:luis_reverseAuction_IEEE_2012} 
    which purposely restrict the rights of contribution and make participants scramble to contribute, have also been used.
    In addition, the willingness-based participant selection~\cite{bib:wang_GPSelector_www}, which selects the participant for requesting based on an estimation of the participant's willingness for sensing task, has been used.
  \item[\dltxt Reward (F4):]
    In this method, participants are compensated for their contribution through monetary 
    and non-monetary incentives~\cite{bib:Gao_incentive_2015, bib:ogie_incentiveMechanisms_springer_2016}.
    The monetary incentive encourages participants to contribute to the system by giving money-related value such as in-app currency/points, discount coupons, and gifts. Participants can directly get explicit value as a consideration for their contribution. It often marks high effectiveness, however, there are problems related to sustainability of system management~\cite{bib:arakawa_gamification_2016}. To reduce costs for rewards, the Auction mechanism~\cite{bib:ioannis_monetaryincentives_taylor2012}, and the following non-monetary incentive mechanism can be used.
    The non-monetary incentive gives \emph{experience} (a kind of fun) as compensation for participant's contribution~\cite{bib:zichermann_oreilly_2011, bib:deterding_mindtrek_2011, bib:groh_ulm_2012}. There are several kinds of experiences such as interaction with other participants, and gamification. The interaction with other participants induces social facilitation effects that stimulate the participant to contribute more actively~\cite{bib:wang_mobilecrowdsourcing_willey2016} for getting praise from others, e.g., getting more \emph{like}, more \emph{comments}. Gamification is the mechanism which introduces game elements into a conventional system, and it has been shown to contribute to motivation of participants, and reduction of monetary rewards~\cite{bib:ueyama_gamification_2014}.
  \item[\dltxt Feedback (F5):]
    This method urges behavior by providing feedback such as a visualization of participants' contribution on a map, graph or timeline.
    Sometimes the contributions of other participants are also included in the visualizations, and sometimes they are excluded.
\end{description}

MinaQn~\cite{bib:sakamura_minaqn_ubicomp_2015} uses a recruitment mechanism to grant participants the right to participate in urban planning (contribution to society) as non-monetary incentive.
In addition, the platform also visualizes a summary of the participant's contribution to increase their willingness to continue contributing.

Medusa~\cite{bib:ra_medusa_mobisys_2012} is a platform which utilizes monetary incentives effectively.
Medusa can acquire participants by using recruitment, which provides money as compensation, and through audition.
Furthermore, to retain participants, Medusa adopts the concept of reverse incentive (obligation/responsibility of executing tasks), where workers pay to organizers for the privilege of performing the task. This can help prevent participants from quitting the system in the middle of sensing tasks.

GP-Selector~\cite{bib:wang_GPSelector_www} is a platform that employs a participant selection algorithm by various conditions suitable for sensing tasks. It includes location-based constraints (e.g., geofences), capability-based constraints (e.g., available sensor), and willingness prediction.

In these conventional platforms, functions to motivate participants have a number of shortcomings.
For example, with Request functions (F3), in order to increase the number of successful requests, it is necessary to consider the notification timing and the target participants, but this is not supported. Similarly, with Reward functions (F4), we need to consider not only monetary incentives, but also \emph{Gamification}.
In this paper, we consider how to design motivating functions which incorporate the concepts of interruption through notification and gamification.

\subsubsection{Processing functions}

In general, organizers intend to analyze or visualize urban environmental data collected with PMS.
Hence, PMS platforms must support easy and quick access to this data.
In the processing functions, the organizer defines methods of data processing to be used.
The following functions are implemented in conventional platforms:

\begin{description}
  \item[\dltxt Data editing (F6):]
    This involves data cleansing and labeling, as pre-processing for detail data analysis.
  \item[\dltxt Data browsing (F7):]
    This involves monitoring the status of data collection and visualizing the collected data.
  \item[\dltxt Data export (F8):]
    This involves exporting of collected data for more detail analysis and/or visualization with third-party tools.
    Various exporting format types are supported depending on the purpose such as CSV, JSON, XML and RDB.
\end{description}

Funf~\cite{bib:funf_social_fMRI_Percom_2011,bib:funf_org} and OpenDataKit~\cite{bib:brunette_opendatakit_hotmobile_2013} are designed as platforms that mainly focus on data cleansing and visualizing, as well as exporting.
Additionally, these platforms support data processing in a variety of environments such as in the cloud and on the smartphones used for data collection (endpoint devices).
Thanks to the various processing functions in the platforms, many research projects in the world utilize them.
Although most of them also support basic functions, there are differences in functions supported by other conventional platforms~\cite{bib:ferreira_aware_frontiers_2015, bib:sakamura_minaqn_ubicomp_2015, bib:tangmunarunkit_ohmage_acmTran_2015}.

In this paper, we implement all functions (F6--F8) as in Funf~\cite{bib:funf_social_fMRI_Percom_2011,bib:funf_org} and
OpenDataKit~\cite{bib:brunette_opendatakit_hotmobile_2013}.
During implementation,
we considered how to make the required technical skills for organizers and participants lower.
Specific skills needed to operate existing platforms are described in the section below.

\ifwithfigure

\tabl[t]{Overview of platform skill requirements.}{tab:skill-requirement-overview}{c|cc|cc}{\hline\hline
   \multirow{4}{5em}{\centering Platforms}
      & \multicolumn{4}{c}{Skill requirements}  \\ \cline{2-5}
        & \multicolumn{2}{c|}{Organizer skills}  & \multicolumn{2}{c}{Participant skills} \\ \cline{2-5}
          & \hw{6.5}{R1}  & \hw{7}{R2}   & \hw{7}{R3}   & \hw{7}{R4}      \\[-2mm]
                & \small Development
                & \small App distribution
                & \small App/Func. management
                & \small Data processing     \\ \hline
    AWARE~\cite{bib:ferreira_aware_frontiers_2015}
        & As needed\footnotemark[1]       & -       & Required  & -     \\
    Sensus~\cite{bib:haoyi_sensus_ubicomp_2016}
        & As needed\footnotemark[2]        & -        & -        & -     \\
    Medusa~\cite{bib:ra_medusa_mobisys_2012}
        & Required  & Required  & -       & -     \\ \hline
    Funf~\cite{bib:funf_social_fMRI_Percom_2011}
        & -       & -       & Required  & Required  \\ 
    MinaQn\footnotemark[3]~\cite{bib:sakamura_minaqn_ubicomp_2015}         & -       & -  & -       & -     \\
    KOKOPIN app\footnotemark[3]~\cite{bib:mishima_kokopin_maed_2013}
        & -            & -        & -        & -     \\ \hline
    Ohmage~\cite{bib:tangmunarunkit_ohmage_acmTran_2015}
        & -       & -       & Required  & -     \\
    OpenDataKit~\cite{bib:brunette_opendatakit_hotmobile_2013}
        & As needed\footnotemark[1] & Required  & Required  & Required  \\ 
    GP-Selector~\cite{bib:wang_GPSelector_www} 
        & As needed\footnotemark[4]   & -     & -     & -  \\ \hline
    ParmoSense 
        & -   & -     & -     & -     \\\hline
}{
\scriptsize
\footnotemark[1] Development by programming is needed for extending functions. \hspace{0.5em}
\footnotemark[2] Database server is required for data collection.\\[0.5mm]
\footnotemark[3] These platforms assume the use of ordinary citizens or administrative officers, hence, technical skills are not required.\\[0.5mm]
\footnotemark[4] XML-based script coding is needed for making complex participant selection constraints.
}{0mm}

\fi

\subsection{Required skills for operation and use}

In PMS, it is assumed that the organizer may be from a non-technical profession/background, e.g.,  they may be an administrative officer or urban planner, and ordinary citizens participate in the data collection.
Thus, it is necessary to re-consider the skills required by the PMS platforms for the organizers and participants.

\tabref{tab:skill-requirement-overview} shows skill requirements for each conventional platform.
Development skills (R1) and App distribution skills (R2)  are required for organizers, where:

\begin{description}
  \item[\dltxt Development skills (R1):]
        Skills to develop the urban sensing system for the specific purpose.
  \item[\dltxt App distribution skills (R2):]
        Skills to distribute client applications to participants' smartphone.
\end{description}

AWARE~\cite{bib:ferreira_aware_frontiers_2015}, Medusa~\cite{bib:ra_medusa_mobisys_2012}, and OpenDataKit~\cite{bib:brunette_opendatakit_hotmobile_2013} have high extendability like a framework, but a high-level of programming skill is required for system construction (R1).
In addition, most of the systems that require system development also require organizers to have the skills necessary to release applications on official stores such as Google Play and AppStore (R2).
With web-based platforms such as MinaQn~\cite{bib:sakamura_minaqn_ubicomp_2015} on the other hand, deploying the applications is quite easy. However, continuously attracting users to the web application is required (R2).

App/Function management skills (R3) and Data processing skills (R4) are other skills that are required from participants, where:

\begin{description}
  \item[\dltxt App/Func. management skills (R3):]
        Managing applications and functions such as  installation of applications and setting of functions to accomplish sensing tasks.
  \item[\dltxt Data processing skills (R4):]
        Processing data collected using the participant's device before uploading.
\end{description}

Since AWARE~\cite{bib:ferreira_aware_frontiers_2015} and
Ohmage~\cite{bib:tangmunarunkit_ohmage_acmTran_2015},
OpenDataKit~\cite{bib:brunette_opendatakit_hotmobile_2013} have many functions as a platform, the functions are divided into multiple applications and provided to participants.
Also, Funf~\cite{bib:funf_social_fMRI_Percom_2011} requires participants themselves to set up the sensors (e.g., sensing interval) for each smartphone.
Therefore, to construct the sensing environment that the organizer intended, knowledge and skills related to applications and functions are required for participants (R3).

Funf~\cite{bib:funf_social_fMRI_Percom_2011} and OpenDataKit~\cite{ bib:brunette_opendatakit_hotmobile_2013} adopt a mechanism where they ask participants to perform data processing and cleansing.
Such processing requires knowledge and skill to judge whether data is good or bad, so the burden on participants is substantial (R4).

Overall, conventional platforms require various skills for both organizers and participants to construct and operate the system.
To realize PMS systems that can be used by non-technical people, these problems must be addressed.
In this study, we propose a new platform to resolve these problems.

\section{Scenario-based participatory mobile sensing platform}
\label{sec:architecture}

We design and implement a novel participatory urban sensing platform, \emph{ParmoSense}, to solve the problems mentioned in \emph{Related work} section. 
ParmoSense aims to solve the following key challenges:

\dlist{
  \item[\normalfont C1: ] Limited support of essential functions needed for PMS systems (listed in \tabref{tab:funcs-overview})
  \item[\normalfont C2: ] Difficulty of system construction and operation
}

ParmoSense allows organizers and participants to operate or contribute to PMS systems without complex procedures or technical skills.

\subsection{ParmoSense Basic Principles}
\label{parmprinciples}

ParmoSense is based on the following three basic principles.

\dlist{
\item[Principle \#1 Modularized functions]\dlbr{0mm}
    ParmoSense must achieve two contradictory requirements, easiness of system construction and diversity of available functions.
    Therefore, we employ the idea of modularizing functions inspired by existing research~\cite{bib:ferreira_aware_frontiers_2015, bib:tangmunarunkit_ohmage_acmTran_2015}.
    ParmoSense provides the sensing, motivating and processing functions in \tabref{tab:funcs-overview}), and these can be combined to form a PMS system. 
\item[Principle \#2 Standardized PMS system]\dlbr{0mm}
    Conventional platforms require a high level of technical skills, e.g., programming skills for organizers to customize the platform for a specific sensing task.
    In order to deal with multiple purposes flexibly, ParmoSense is composed as a combination of modularized functions as well as the detailed settings of each function. We unify the combination and settings as a \emph{scenario}.
    A \emph{Scenario} contains such information as the scenario name, description, sensing targets, sensing area, period, motivation method, etc. Through GUI-tool in ParmoSense, the organizer can generate the scenario easily. Based on the created scenario, ParmoSense automatically configures both a server system and a client application by distributing necessary information. 
    Since scenarios can be created using any combination of functions and settings, logically, any kinds of PMS system can be built with ParmoSense. Another important advantage of the scenario-based system is that users can participate in various PMS projects through one client application, whereas in the past, each PMS project required a dedicated application.
\item[Principle \#3 Customizable motivation engine]\dlbr{0mm}
    The most effective way of motivating participants depends on the purpose of sensing.
    To realize sustainable urban environmental sensing, ParmoSense has a \emph{Motivation Engine} with a variety of motivation algorithms.
    The Motivation Engine provides the following functions for motivating participants:
    \ilist{
      \item \textbf{Motivation based on the behavior of an individual:}\dlbr{0mm}
          Granting incentives according to contribution, Visualization of contribution
      \item \textbf{Motivation for all participants, regardless of contribution:}\dlbr{0mm}
          Providing competition mechanisms such as rankings, Sharing of experiences among participants
    }
    Additionally, by considering temporal/spatial information, e.g., the current time and the current position of the participant, it is possible to control the actuation timings of these functions.
    The optimal motivating algorithm according to the purpose of the organizer can be incorporated into the PMS system by combining and customizing these functions.
}

\fig{width=0.75\columnwidth}{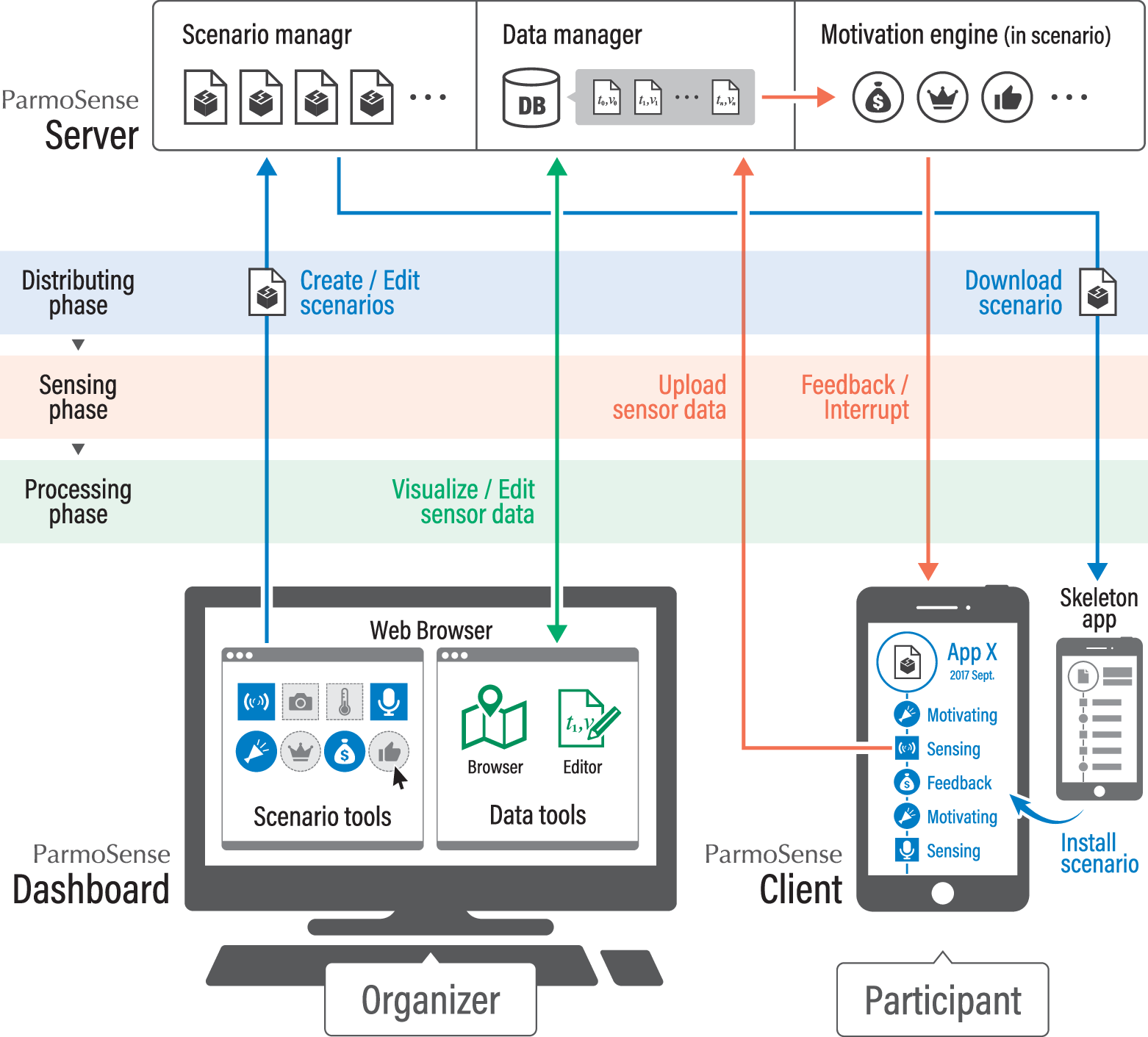}{\textbf{Design concept of ParmoSense.}}{fig:design_concept}{}

\subsection{ParmoSense system overview}

The architectural design of ParmoSense is shown in \figref{fig:design_concept}.
ParmoSense consists of three parts that have the following roles:

\begin{description}
  \item[\dltxt \ParmoSenseDashboard:]\dlbr{0mm}
        A web application for organizers that can be used to create and distribute scenarios of PMS systems, and to process collected data.
  \item[\dltxt \ParmoSenseClient:]\dlbr{0mm}
        A client application for participants that can run various scenarios.
        By downloading and installing a scenario, it behaves as the corresponding sensing application.
  \item[\dltxt \ParmoSenseServer:]\dlbr{0mm}
        A central system for integrated management of the scenarios created by organizers and automatically constructing a virtual server system for each scenario.
        It can collect sensing data from \ParmoSenseClient and generate feedback based on the analysis of the collected data.
\end{description}

Arrows in \figref{fig:design_concept} show contents transfered between organizer/participant and \ParmoSenseServer for each operation phase of PMS.
Each phase is defined as follows:

{
\begin{enumerate}
  \item \textbf{Distributing Phase:}\ \ 
      \parbox[t][][t]{34.5em}{
        The phase for distributing the scenario created by the organizer to the participants via \ParmoSenseServer.
      }\\[1mm]
  \item \textbf{Sensing Phase:}\hspace{2.05em}
      \parbox[t][][t]{34.5em}{
        The phase for giving feedback (e.g., reward) to the contribution of the participant such as uploading sensing data by participants. 
      }\\[1mm]
  \item \textbf{Processing Phase:}\hspace{0.71em}
      The phase for editing, cleansing and visualizing the collected data.
\end{enumerate}
}

Since ParmoSense is based on \textit{Principles \#1, \#2} mentioned above,
organizers can distribute sensing applications by simply exchanging scenarios with participants in Distributing Phase.
It is therefore not necessary for each participant to manage the applications by himself.
Furthermore, thanks to \textit{Principle \#3}, 
in Sensing Phase,
the feedback for motivating participants is created by the Motivation Engine using the collected data,
and automatically provided to participants.

\fig{width=\columnwidth}{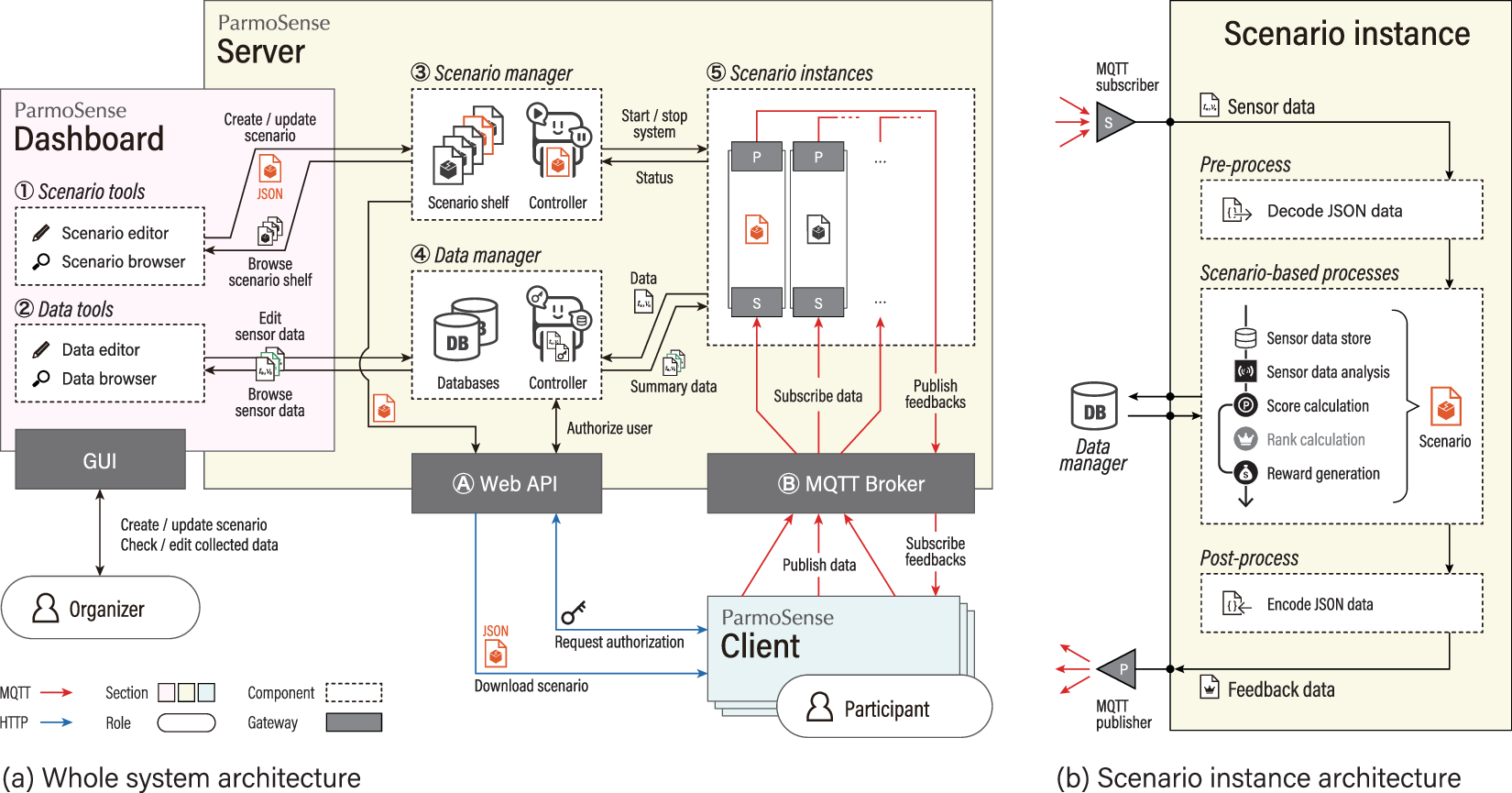}{\textbf{System architecture of \ParmoSense.} (a) ParmoSense consists of three components: Dashboard, Server, and Client; (b) Scenario instances are built for each PMS scenario.}{fig:system-architecture}{}

\subsection{ParmoSense system architecture}
\label{subsec:parmosense-system-architecture}

The concrete system configuration of ParmoSense is shown in \figref{fig:system-architecture}.
In the following subsections, we will describe the \ParmoSenseDashboard used by organizer, the \ParmoSenseClient used by participants, and the \ParmoSenseServer in more detail.

\fig{width=\columnwidth}{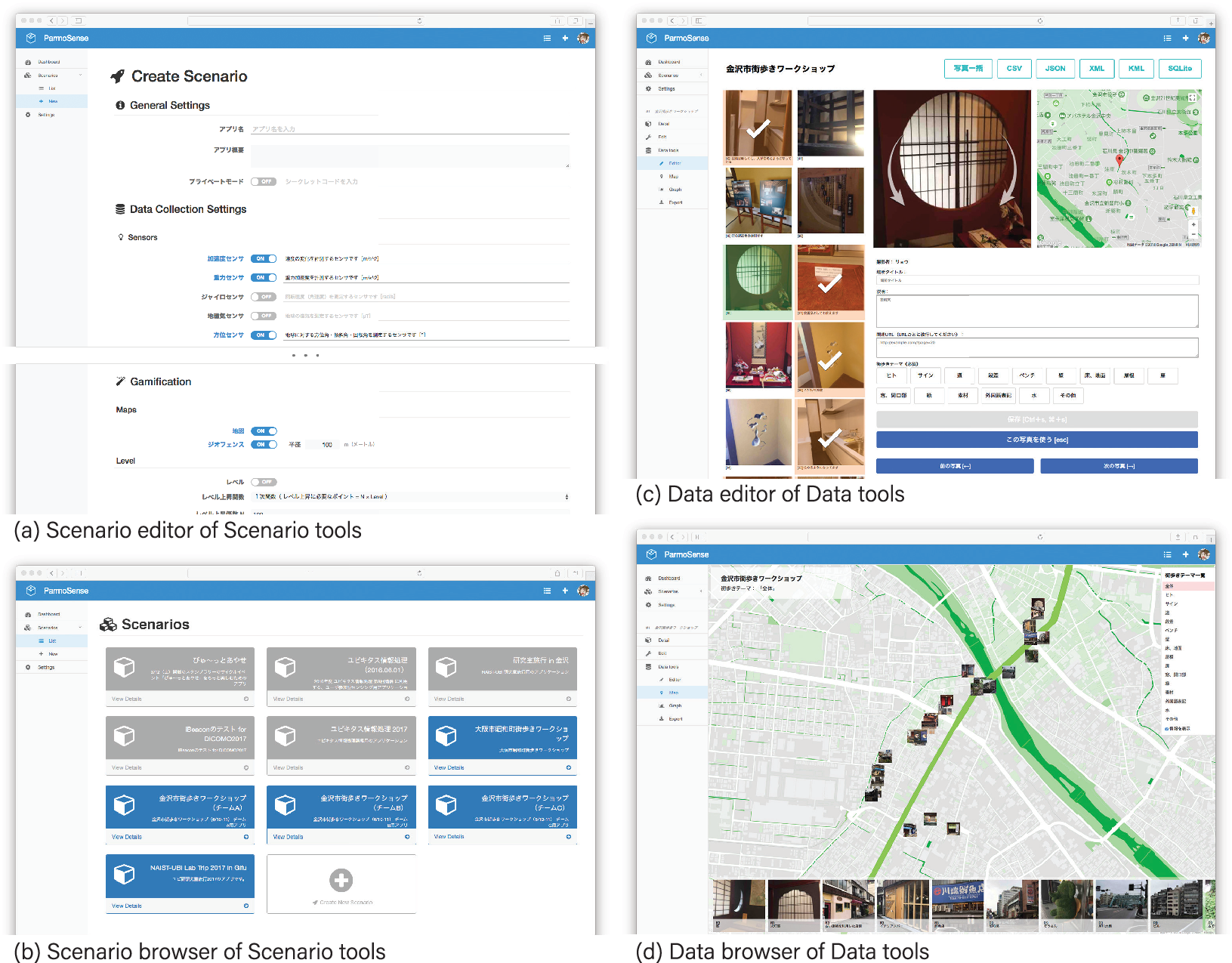}{\textbf{User interface of \ParmoSenseDashboard.} (a),~(b)~\ScenarioTools interface as shown in \figref{fig:system-architecture} (a) \cnum{1}; (c),~(d)~\DataTools interface as shown in \figref{fig:system-architecture} (a) \cnum{2}.}{fig:dashboard-UI}{}

\subsubsection{\ParmoSenseDashboard}

An organizer carries out every operation, e.g., management of a PMS scenario, processing of collected data on the \ParmoSenseDashboard, a web application. It consists of \emph{\ScenarioTools} and \emph{\DataTools} shown in \figref{fig:system-architecture} (a) \cnum{1} and \cnum{2} respectively.

\ScenarioTools (\figref{fig:system-architecture} (a) \cnum{1}) provide many operations such as creating, editing and deleting PMS scenarios, and browsing, activating and stopping scenarios.
\figref{fig:dashboard-UI} (a) shows the user interface for editing scenarios.
The organizer can describe a scenario using the three kinds of functions, sensing, motivating and processing functions, mentioned in \emph{Related work} section, 
without programming, through the GUI (\emph{Principles \#1, \#3\,}).
The scenario defined in \ScenarioTools is automatically converted to JSON format, and transferred between each part of ParmoSense.

When the scenario editing is completed, the virtual server system is automatically built depending on the scenario, and deployed by \emph{\ScenarioManager} (\figref{fig:system-architecture} (a) \cnum{3}).
At the same time, the QR code for downloading the scenario to \ParmoSenseClient is automatically generated.
\figref{fig:dashboard-UI} (b) shows the user interface for browsing/managing the scenario created by organizers.
Scenarios currently in progress/stopped are indicated by blue/gray respectively, and these statuses and the scenario settings can be changed by using the GUI.

Data tools (\figref{fig:system-architecture} (a) \cnum{2}) provide the functions for processing and visualizing data aggregated into \emph{\DataManager} (\figref{fig:system-architecture} (a) \cnum{4}).
The user interface for editing the collected data is shown in \figref{fig:dashboard-UI} (c).
The organizer can improve the quality of the data by editing/excluding inappropriate data from the collected data. 
The organizer can also add labels to the data.
The processed data can be exported in the form of JSON, CSV, RDB etc.
The user interface for visualizing the collected data is shown in \figref{fig:dashboard-UI} (d).
The organizer can check the data in two ways: overlaying them on a geographical map, or sorting them in a time-series order.

\fig{width=\columnwidth}{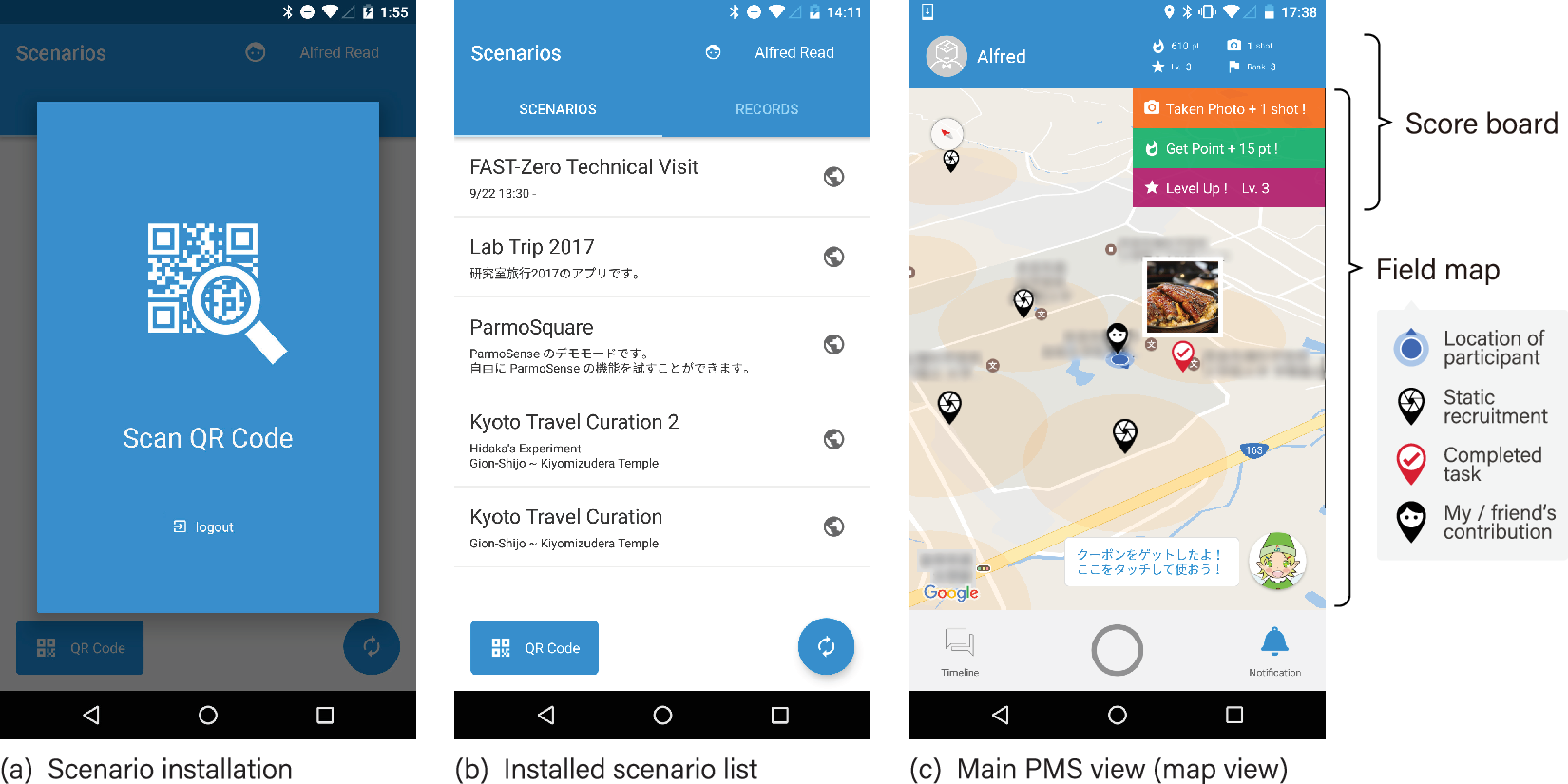}{\textbf{User interface of \ParmoSenseClient.} (a) Participants can install a PMS scenario via scanning the QR code; (b) Installed scenarios are listed up for switching scenarios; (c) PMS tasks and participants' contributions are shown on a map view.}{fig:client-UI}{}

\subsubsection{\ParmoSenseClient}

The participant performs all sensing tasks of ParmoSense through the \ParmoSenseClient smartphone application.
\ParmoSenseClient runs on smartphones with Android OS or iOS, and it can be installed from general application stores (Google Play, AppStore).
Since the behavior of PMS system on ParmoSense is defined by a scenario ({\it Principle \#2\,}),
it can behave as various sensing applications by installing scenarios to \ParmoSenseClient.

\figref{fig:client-UI} (a) and (b) shows the user interface for the scenario installation.
The participant performs the following steps in order to install the application:

\begin{enumerate}
    \item Log in on \ParmoSenseClient via Google Authentication.
    \item Scan the scenario QR code by using their smartphone camera (\figref{fig:client-UI} (a)). An organizer can get QR codes of scenarios from \ParmoSenseDashboard, and print it out for participants.
    \item The participant confirms participation in sensing.
\end{enumerate}

This procedure makes it easy to install scenarios.
All these are done via the \emph{Web API} shown in \figref{fig:system-architecture} (a) \cnum{A}.
Participants can participate in multiple scenarios.
The scenarios that have been installed and performed in the past are listed, as shown in \figref{fig:client-UI} (b).
This makes it easy for a participant to join the same scenario again.

\figref{fig:client-UI} (c) shows an example of the \ParmoSenseClient interface after installing the scenario and participating in sensing tasks.
Requests of static sensing tasks are shown as pins on the map, and the participants can carry out the task at this place and acquire the reward accordingly.
The user score is displayed in the upper right area.
It is a means of providing feedback to the participant through gamification and visualization of contributions.
The feedback and the execution status of tasks are reflected in real time according to the participant's and other participants' actions via \emph{MQTT Broker} shown in \figref{fig:system-architecture} (a) \cnum{B}.
ParmoSense realizes many-to-many and real-time communication by adopting MQTT (MQ Telemetry Transport)~\cite{bib:mqtt_org} as a communication protocol.

\fig{width=\columnwidth}{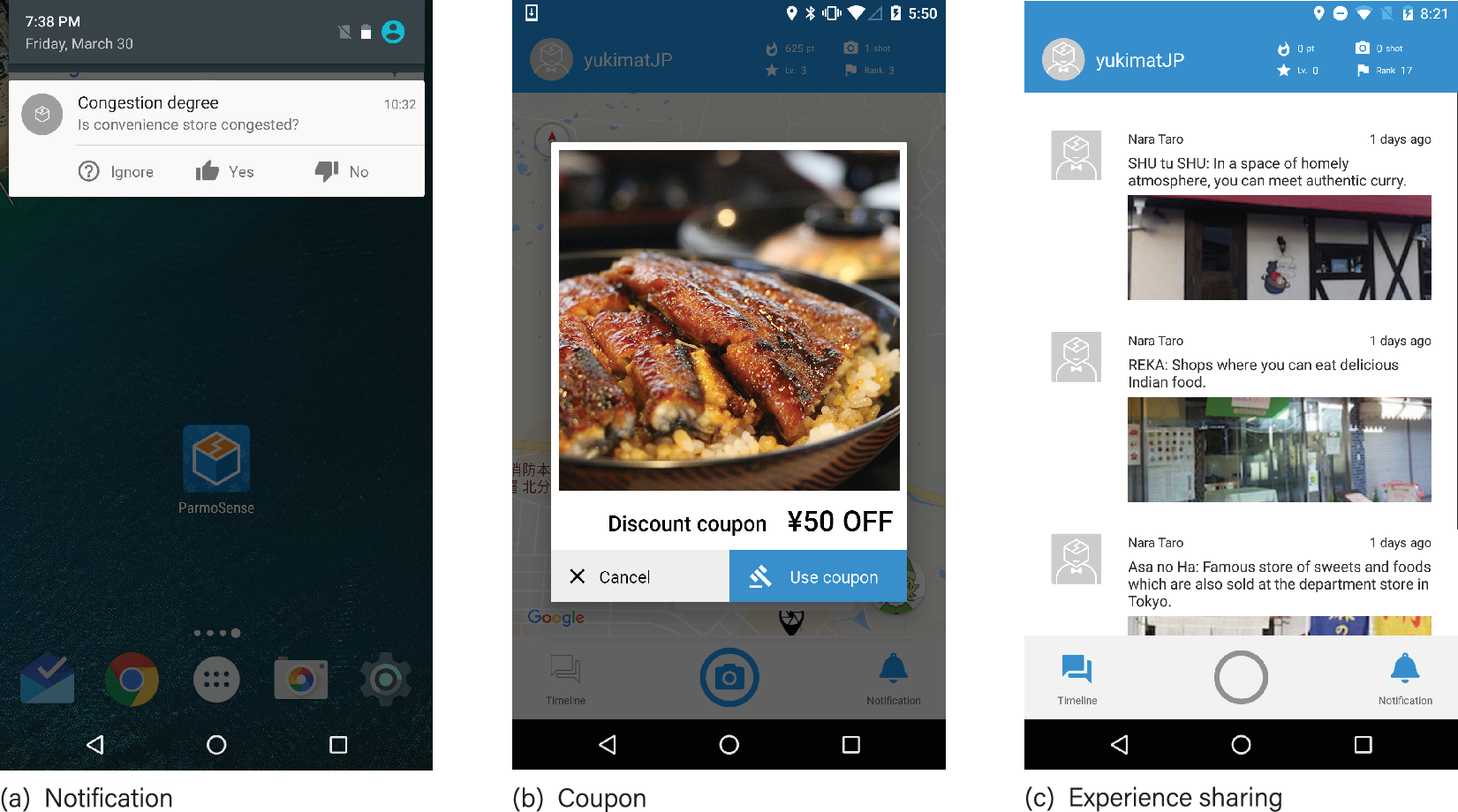}{\textbf{Examples of \MotivatingFuncs.} (a) \ParmoSenseClient can send a notification to participants for requesting a PMS task; (b) ParmoSense can provide coupon as monetary incentive; (c) Other participants' contributions are shared as a timeline.}{fig:client-motivating-funcs}{}

\subsubsection{\ParmoSenseServer}

\ParmoSenseServer consists of three parts: \emph{\ScenarioManager}, \emph{\DataManager} and \emph{\InternalSystems} shown in \figref{fig:system-architecture} (a) \cnum{3}, \cnum{4} and \cnum{5} respectively.

\ScenarioManager plays the role of storing the scenario created on \ParmoSenseDashboard, and constructing and managing \InternalSystem in accordance with the scenario.
The \InternalSystem is an instance executed on a server which communicates with a \ParmoSenseClient.
By automatically constructing this \InternalSystem for each scenario and forming \InternalSystems, various scenarios in ParmoSense can be created without programming.
\ScenarioManager monitors the operation status of the \InternalSystem, and the organizer can start or stop the operation.
It also detects unexpected troubles in the \InternalSystem, and it stops or restarts them.
The data collected by participants is aggregated in \DataManager, and this data is used for calculating the participants' score, visualizing on the map and so on.

\figref{fig:system-architecture} (b) shows the mechanism of the \InternalSystem.
\ScenarioManager builds the \InternalSystem by incorporating the module program of corresponding functions (Sensing Functions, Motivating Functions, Processing Functions) based on the scenario that an organizer created.
\InternalSystem communicates with \ParmoSenseClient via MQTT as described above.
If a participant sends (publish) the sensing data, the corresponding \InternalSystem receives (subscribe) the data collected by participant's smartphone sensors, and processes using modularized functions (e.g., analysis of data, calculation of ranking) which are described in the scenario.
According to these results, response data is generated and published to all participants who should be informed.

\subsection{Functions support}

ParmoSense comprehensively supports functions investigated in \tabref{tab:funcs-overview}.
In this section, we outline the support status of each function.

\boldTitle{Implicit-sensing (F1)}
ParmoSense supports data collection from sensors embedded in smartphones.
The types of supported sensors are shown below:

\begin{itemize}
  \item Position sensor (e.g., GPS)
  \item Environmental sensors (e.g., light sensor, barometer)
  \item Inertial sensor (e.g., accelerometer, gyroscope)
  \item External sensor device (heart-rate sensor)
\end{itemize}

It also supports data collection of BLE scan logs of peripheral devices (e.g., iBeacon, other smartphones).
For all these sensors, detailed configuration such as measurement interval and enabling/disabling of background measurement can be set on \ScenarioEditor of \ParmoSenseDashboard by the organizer.

\boldTitle{Explicit-sensing (F2)}
ParmoSense supports various kinds of data collection methods that cannot be collected by sensors embedded in a smartphone.
One of them is photo uploading.
When taking photos and uploading them, participants can provide additional data such as explanatory texts of the photo taken,  GPS position, and other data obtained by Implicit-sensing.
Questionnaires are also provided. Different question types such as binary questions (YES/NO question), 
multiple-choice questions (up to 4 options), and questions that require photo uploads and explanatory text, are supported.
These questions can be chained together to support a step-by-step questionnaire.

\boldTitle{Static/Dynamic request (F3)}
ParmoSense supports both static and dynamic requests for soliciting contributions on sensing tasks.
In PMS for urban environments, there are many requests based on geographical information.
Static requests place each task as a checkpoint on a map as shown in \figref{fig:client-UI} (c), which allows participants to easily find the tasks to be performed.
For dynamic requests, we support informing participants of task requests through notifications as shown in \figref{fig:client-motivating-funcs} (a).
Organizers can generate and request tasks from specific participants, for example, by setting that any person who is detected entering a certain area is notified. Location information can be obtained using region monitoring technology such as Geofence and iBeacon.

\boldTitle{Monetary/Non-monetary reward (F4)}
ParmoSense supports both monetary and non-monetary rewards to compensate participants for their contributions.
The monetary rewards supported are discount coupons that can be used at restaurants, cafe and so on, as shown in \figref{fig:client-motivating-funcs} (b).
Since the discount coupon is linked with purchasing behavior, it is effective for motivating participation in scenarios such as sightseeing. 
For non-monetary rewards, ParmoSense supports the gamification mechanism.
This includes awarding points for a contribution, competition mechanisms such as comparing a participant's degree of contribution to that of other participants, and virtual level-up elements by repeating the contribution.
Also, depending on the demand for data collection, monetary/non-monetary rewards can be biased. For example, the expensive reward can be set for a low upload rate task or high priority task.

\boldTitle{Feedbacks by visualization  (F5)}
ParmoSense also supports feedback not included in game features, for example, visualization of own contributions, and data/experience sharing.
Visualization of own contributions is of two types:  (i)  plotting the pins of contributions on the map and (ii) scoring contributions with the non-monetary rewards detailed above.
In addition, a data sharing function to share/visualize data such as participants' experiences and what different participants viewed, heard, or sensed (environmental conditions) is provided.
For example, ParmoSense can plot sensing data uploaded by other participants on the map and share them on a timeline as shown in \figref{fig:client-motivating-funcs} (c).

\boldTitle{Editing of collected data (F6)}
ParmoSense helps organizers to pre-process collected data before detailed analysis.
For example, it provides functions such as cleansing unnecessary data, selecting data to be used, and labeling the data.
Also, since these data processing functions are designed to protect the original data, it can be restored at any time.

\boldTitle{Browsing of collected data (F7)}
ParmoSense provides visualization tools for instant and easy confirmation of the collected data.
There are many types of tools such as a tool for plotting data collected by all or each participant on the map, and a tool for displaying all data as a list.
These tools can be used at any time even while sensing tasks are in progress.

\boldTitle{Export of collected data (F8)}
ParmoSense supports various data output formats.
Data analysis data can be output in CSV, JSON, XML, RDB (SQLite), etc., so that analysis can be started immediately.
Moreover, when exporting to a third-party visualization tools (e.g., Open Street Map~\cite{bib:openstreetmap_org}, Google Earth~\cite{bib:google_earth}, Cesium~\cite{bib:cesiumjs_org}), it is possible to output in KML (\url{https://developers.google.com/kml/}) or GPX (\url{http://www.topografix.com/gpx.asp}).

\section{Evaluation}
\label{sec:evaluation}

ParmoSense has tackled two challenges to be solved: ``limited support of essential functions in existing PMS platforms (C1),'' and ``difficulty of system construction and operation (C2).'' In this section, we use the \emph{easiness of system construction and operation} and the \emph{number of functions provided} as the metrics, and we evaluate the performance of ParmoSense compared with conventional platforms.

\tabl[t]{Preparation Workload for Each Task.}{tab:workload_time}{c|l|l|l}{\hline\hline
      & Contents of work & Preparation Workload & Done by \\\hline
        $w_1$ & development (programming)   & 0, 8\footnotemark[1] 
            & \multirow{3}{5em}{\organizer} \\
        $w_2$ & development (GUI editing)   & 0, 4\footnotemark[5] & \\
        $w_3$ & publishing to app store     & 0, 8\footnotemark[2] & \\\hline
        $w_4$ & installing application      & 0, 1\footnotemark[3], 2\footnotemark[4] 
            & \multirow{2}{5em}{\participant} \\
        $w_5$ & configuring functions       & 0, 2\footnotemark[5] & \\\hline
}{
\scriptsize
\footnotemark[1]~calculated with LOC of source code.
\footnotemark[2]~calculated based on time for becoming available in general stores.
\footnotemark[3]~installation of single application (reference time).
\footnotemark[4]~installation of multiple application.
\footnotemark[5]~calculated by comparing with $w_1$, $w_3$ and $w_4$ relatively.
}{0em}

\subsection{Workload for system operation}

We evaluated the difference in the development cost and the operation cost of ParmoSense compared to conventional platforms~\cite{bib:ferreira_aware_frontiers_2015, bib:haoyi_sensus_ubicomp_2016, bib:ra_medusa_mobisys_2012, bib:funf_social_fMRI_Percom_2011, bib:sakamura_minaqn_ubicomp_2015, bib:mishima_kokopin_maed_2013, bib:tangmunarunkit_ohmage_acmTran_2015, bib:brunette_opendatakit_hotmobile_2013, bib:wang_GPSelector_www}.
We used the workload cost for starting operation of the system (Preparation Workload) as the metric of comparison.
The definition of Preparation Workload is as follows:

\begin{itembox}[l]{Preparation Workload ($W$) }
Preparation Workload ($W$) is the total time required from the start of developing the PMS system to the start of sensing tasks, which is calculated by the sum of sub tasks (\eqnref{eqn:workload-score}). 
\eqn{
    &&T = \sum_{i=1}^{5} w_i\hspace{4em} (0\leq W)\hspace{2em} \label{eqn:workload-score}
}
Where $w_1 \sim w_5$ is the relative estimated time required for each task, as shown in \tabref{tab:workload_time}. As reference, we defined the time required to install one application from a general application store, e.g., Google Play or AppStore as $w_x = 1$.
\end{itembox}

The estimated Preparation Workload on each platform is shown in \tabref{tab:evaluation-metrics} and the x-axis of \figref{fig:comparison_graph}.
The filled circle ({\large $\bullet$}) represents the case where the simplest system on each platform was constructed.
AWARE~\cite{bib:ferreira_aware_frontiers_2015},
Medusa~\cite{bib:ra_medusa_mobisys_2012}, 
OpenDataKit~\cite{bib:brunette_opendatakit_hotmobile_2013} and
GP-Selector~\cite{bib:wang_GPSelector_www} can be extended by programming as necessary.
However, $w_1$ will increase linearly with the development of functionality.

Therefore, \ParmoSense belongs to the group which needs low Preparation Workload, and can be operated as easily as other platforms in the same group.
Also, conventional platforms~\cite{bib:ferreira_aware_frontiers_2015, bib:ra_medusa_mobisys_2012, bib:brunette_opendatakit_hotmobile_2013, bib:wang_GPSelector_www} suffer from the problem where the time required for expansion tends to be long when trying to extend the systems' functionalities.
In contrast, \ParmoSense is designed to cover the necessary functions in advance, and thus, although Preparation Workload required is equivalent to other platforms, it achieves higher functionality.

\subsection{Variety of functions}

We evaluated the diversity of functions that ParmoSense provides, compared to conventional platforms~\cite{bib:ferreira_aware_frontiers_2015, bib:haoyi_sensus_ubicomp_2016, bib:ra_medusa_mobisys_2012, bib:funf_social_fMRI_Percom_2011, bib:sakamura_minaqn_ubicomp_2015, bib:mishima_kokopin_maed_2013, bib:tangmunarunkit_ohmage_acmTran_2015, bib:brunette_opendatakit_hotmobile_2013, bib:wang_GPSelector_www}.
As a metric for comparison, we use the following fulfillment status of functions (Function Score):

\begin{itembox}[l]{Function Score ($S$)}
\tabref{tab:funcs-overview} shows all the functions supported by each of the existing PMS platforms.
The score of each function $s_1 \sim s_8$ is determined by the implementation status of each function (F1 $\sim$ F8) of \tabref{tab:funcs-overview}, and Function Score ($S$) is calculated by the sum of them (\eqnref{eqn:func-score}).
\eqn{
      &&S = \sum_{i=1}^{8} s_i\hspace{4em} (0\leq S \leq 8)\hspace{2em} \label{eqn:func-score}\\
      &&s_i = \left\{\begin{array}{cll}
              1    && \textrm{if\ \ F}[i]=\checkmark \\[-1.5mm]
              0.5  && \textrm{if\ \ F}[i]=\triangle \\[-1.5mm]
              0    && \textrm{otherwise}
              \end{array}\right. \nonumber
}
\end{itembox}

The estimated Function Score on each platform is shown in \tabref{tab:evaluation-metrics} and the y-axis of \figref{fig:comparison_graph}.
While conventional platforms have moderate scores ($S=4 \pm 1$), ParmoSense has a higher score ($S=8$) due to comprehensively supporting all the functions provided in existing platforms, and incorporating motivating functions, which are missing on all existing platforms.
Furthermore, the advantage of ParmoSense will be even higher if we consider the effect brought by the combination of these functions. For example, ParmoSense can provide combinations of multiple motivating functions, such as gamification, rewards, and interruptions, which can not be provided on other existing platforms, and which may be more effective than using a single motivating strategy/function, since they cater for more participant preferences.

\tabl[t]{Breakdown of Preparation Workload ($W$) and Function Score ($S$).}
{tab:evaluation-metrics}{c|ccc|cc|c|c}{\hline\hline
   \multirow{2}{5em}{\centering Platforms} &
    \multicolumn{6}{c|}{Preparation Workload} &
    Function Score \\\cline{2-8}
    & $w_1$   & $w_2$  & $w_3$  & $w_4$   & $w_5$   & $W$   & $S$ \\ \hline
AWARE~\cite{bib:ferreira_aware_frontiers_2015}
    & -\footnotemark[1] & 4 & -  & 2 & 2 & 8       & 4.5     \\
Sensus~\cite{bib:haoyi_sensus_ubicomp_2016}
    & -                 & 4 & -  & 1 & - & 5       & 3       \\
Medusa~\cite{bib:ra_medusa_mobisys_2012}
    & 8\footnotemark[1] & - & 8  & 1 & - & 17      & 3.5     \\\hline
Funf~\cite{bib:funf_social_fMRI_Percom_2011}
    & -                 & - & -  & 1 & 2 & 3       & 4.5     \\
MinaQn~\cite{bib:sakamura_minaqn_ubicomp_2015}
    & -                 & 4 & -  & - & - & 4       & 3.5     \\
KOKOPIN app~\cite{bib:mishima_kokopin_maed_2013}
    & -                 & 4 & -  & 1 & - & 5       & 4       \\\hline
Ohmage~\cite{bib:tangmunarunkit_ohmage_acmTran_2015}
    & -                 & - & -  & 2 & 2 & 4       & 5       \\
OpenDataKit~\cite{bib:brunette_opendatakit_hotmobile_2013}
    & -\footnotemark[1] & 4 & 8  & 2 & 2 & 16      & 4.5     \\
GP-Selector~\cite{bib:wang_GPSelector_www}
    & -\footnotemark[1] & 4 & -  & 1 & - & 5       & 3.5     \\\hline
ParmoSense 
    & -                 & 4 & -  & 1 & - & 5       & 8       \\\hline
}{
\scriptsize
\footnotemark[1] Additional time is required for developing the new functionalities.
}{1em}

\fig{width=0.7\columnwidth}{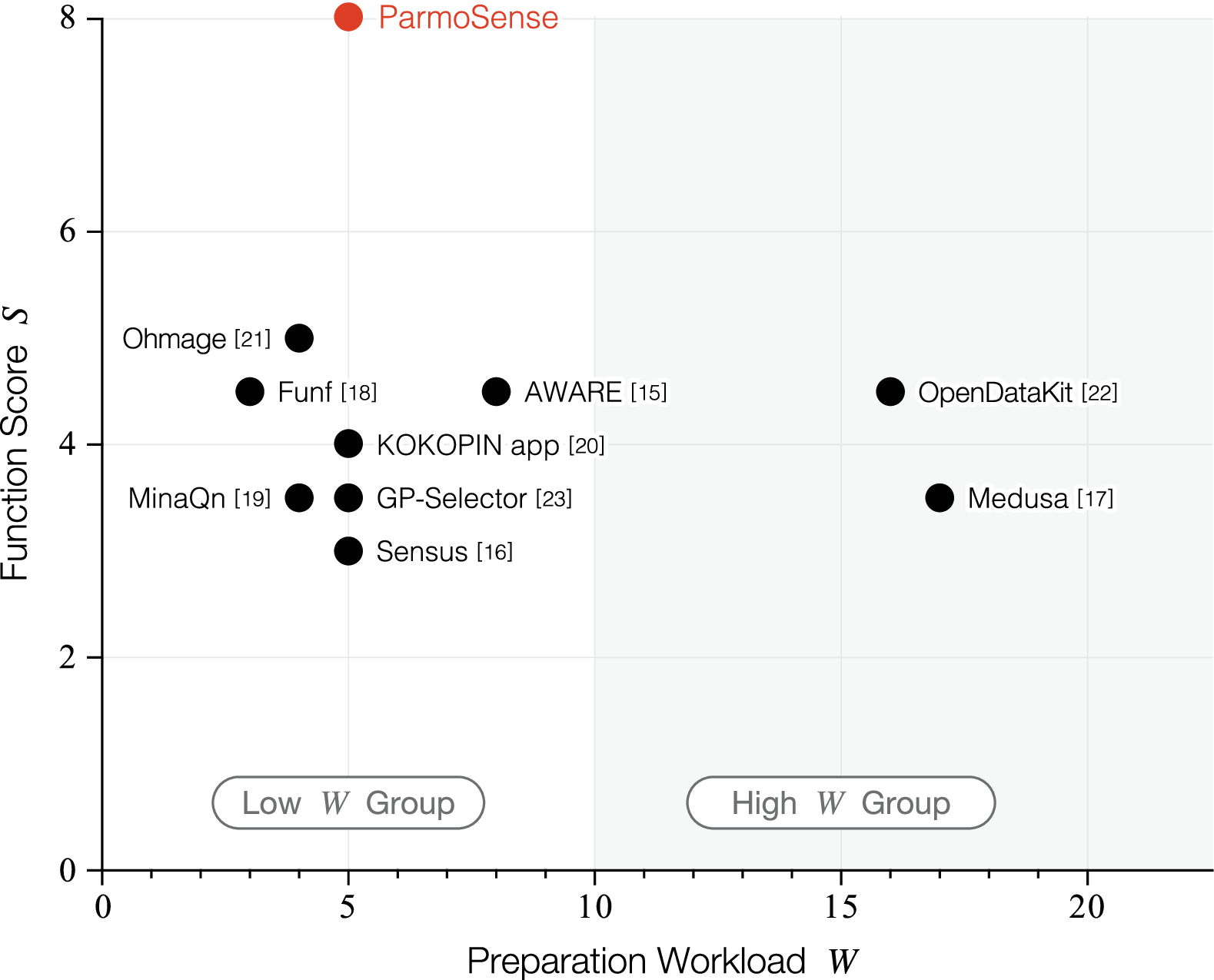}{\textbf{Comparison graph.}}{fig:comparison_graph}{}

\newpage

\section{Case studies}
\label{sec:case_studies}

We conducted 19 case studies with ParmoSense over four years. Prior to the evaluation, we released \ParmoSense to general application stores (Google Play, AppStore). We collaborated with various organizers to design and build 19 scenarios, and then deployed them via the client application. Members of the general public who had downloaded the application acted as participants.

Our aim in doing these case studies was two-fold. First, we wanted to validate that our implementations of the sensing, processing and motivating functions on ParmoSense would work well in real world PMS tasks. Second, we wanted to determine how effective the functions were in motivating participation and in supporting organizers to collect required data and to extract the required information. Such knowledge would allow us to improve the functions, or to recommend additional functions to be included in PMS systems. 
\tabref{tab:case_studies} shows the details of each scenario (start date, period, number of \participants, and embedded \PSfunctions).
In following subsections, we discuss the sufficiency of \ParmoSense functions in each scenario.


\newcommand{\UbiLectA}{S1\xspace}
\newcommand{\Suita}{S2\xspace}
\newcommand{\UbiLectB}{S3\xspace}
\newcommand{\Ikoma}{S4\xspace}
\newcommand{\OCUnivA}{S5\xspace}
\newcommand{\OCUnivB}{S6\xspace}
\newcommand{\UbiLabA}{S7\xspace}
\newcommand{\NICTA}{S8\xspace}
\newcommand{\NICTB}{S9\xspace}
\newcommand{\AIPA}{S10\xspace}
\newcommand{\AIPB}{S11\xspace}
\newcommand{\MThesis}{S12\xspace}
\newcommand{\UbiLectC}{S13\xspace}
\newcommand{\AyaseA}{S14\xspace}
\newcommand{\CICP}{S15\xspace}
\newcommand{\AyaseB}{S16\xspace}
\newcommand{\UbiLabB}{S17\xspace}
\newcommand{\OsakaUnivA}{S18\xspace}
\newcommand{\OsakaUnivB}{S19\xspace}

\ifwithfigure

\begin{sidewaystable}[htbp]
\centering
  \caption{\textbf{Overview of Case Studies.}}
  \label{tab:case_studies}
  \begin{tabular}{c|cc|ccc|ccc|c|rl|c}\multicolumn{13}{c}{}\\[-3mm]\hline\hline
\multirow{4}{3.7em}{\centering \shortstack{Scenario\\No.}}  & \multicolumn{8}{c|}{\PSFunctions}                             & \multirow{4}{5em}{\centering Start date}  & \multicolumn{2}{c|}{\multirow{4}{3.6em}{\centering Period}}     & \multirow{4}{6em}{\centering Number of participants} \\ \cline{2-9}
  & \multicolumn{2}{c|}{\SensingFuncs}      & \multicolumn{3}{c|}{\MotivatingFuncs}           & \multicolumn{3}{c|}{\ProcessingFuncs}         &   &   &   &   \\ \cline{2-9}
  & \hw{3.2}{F1}  & \hw{3.2}{F2}  & \hw{2.6}{F3}  & \hw{2.6}{F4}  & \hw{2.6}{F5}  & \hw{2.6}{F6}  & \hw{2.6}{F7}  & \hw{2.6}{F8}  &   &   &   &   \\[-2.3mm]
  & {\scriptsize Implicit}  & {\scriptsize Explicit}  & {\scriptsize Request} & {\scriptsize Reward}  & {\scriptsize Feedback}  & {\scriptsize Editing} & {\scriptsize Browsing}  & {\scriptsize Export}  &   &   &   &   \\ \hline
\UbiLectA  & \checkmark  & \checkmark  &   & \checkmark  & \checkmark  &   & \checkmark  &   & Feb. 9th, 2016  & 0.5 &\hspace{-1em}  hours & 15  \\
\Suita  & \checkmark  & \checkmark  &  &  & \checkmark & \checkmark & \checkmark & \checkmark & Feb. 17th, 2018 & 1 &\hspace{-1em}  day & 9 \\
\UbiLectB  & \checkmark  & \checkmark  &   & \checkmark  & \checkmark  &   & \checkmark  &   & June 1st, 2016  & 0.5 &\hspace{-1em}  hours & 17  \\
\Ikoma  & \checkmark  & \checkmark  &   &   & \checkmark  & \checkmark  & \checkmark  & \checkmark  & Dec. 17th, 2016 & 2 &\hspace{-1em}  days  & 14  \\
\OCUnivA  & \checkmark  & \checkmark  &   &   & \checkmark  & \checkmark  & \checkmark  & \checkmark  & June 8th, 2017  & 1 &\hspace{-1em}  day & 6 \\
\OCUnivB  & \checkmark  & \checkmark  &   &   & \checkmark  & \checkmark  & \checkmark  & \checkmark  & June 10th, 2017 & 1 &\hspace{-1em}  day & 20  \\ \hline
\UbiLabA  & \checkmark  & \checkmark  &   & \checkmark  & \checkmark  &   & \checkmark  & \checkmark  & June 5th, 2016  & 2 &\hspace{-1em}  days  & 19  \\
\NICTA~\cite{bib:hidaka_curation_smartcities2019}  & \checkmark  & \checkmark  & \checkmark  &   & \checkmark  & \checkmark  & \checkmark  & \checkmark  & Nov. 22nd, 2016 & 3 &\hspace{-1em}  days  & 14  \\
\NICTB~\cite{bib:shltn_gamification_smartcities2020}  & \checkmark  & \checkmark  & \checkmark  & \checkmark  & \checkmark  &   &   &   & Nov. 26th, 2017 & 1 &\hspace{-1em}  day & 30  \\
\AIPA~\cite{bib:shltn_tourism_mobiquitous2020}  & \checkmark  & \checkmark  & \checkmark  & \checkmark  & \checkmark  &   &   &   & Jul. 29th, 2020 & 1 &\hspace{-1em}  day & 10  \\
\AIPB  & \checkmark  & \checkmark  & \checkmark  & \checkmark  & \checkmark  &   &   &   & Oct. 5th -- Nov. 4th, 2020 & 1 &\hspace{-1em}  day/person & 108  \\ \hline
\MThesis  & \checkmark  & \checkmark  & \checkmark  & \checkmark  & \checkmark  &   &   &   & Jan. 21st, 2016 & 10  &\hspace{-1em}  days  & 12  \\
\UbiLectC & \checkmark  & \checkmark  & \checkmark  &   &   &   & \checkmark  &   & Apr. 20th, 2017 & 14  &\hspace{-1em}  days  & 83  \\ \hline
\AyaseA & \checkmark  & \checkmark  & \checkmark  & \checkmark  & \checkmark  &   & \checkmark  & \checkmark  & Mar. 12th, 2016 & 1 &\hspace{-1em}  day & 10  \\
\CICP & \checkmark  & \checkmark  & \checkmark  & \checkmark  & \checkmark  &   & \checkmark  &   & Nov. 13th, 2016 & 1 &\hspace{-1em}  day & 48  \\
\AyaseB & \checkmark  & \checkmark  & \checkmark  & \checkmark  & \checkmark  &   & \checkmark  & \checkmark  & Feb. 25th, 2017 & 1 &\hspace{-1em}  day & 25  \\
\UbiLabB & \checkmark  &   &   &   &   &   &   & \checkmark  & June 11th, 2017 & 2 &\hspace{-1em}  days  & 18  \\
\OsakaUnivA & \checkmark  &   &   &   &   &   &   & \checkmark  & July 24th, 2017 & 4 &\hspace{-1em}  hours & 100 \\
\OsakaUnivB & \checkmark  &   &   &   &   &   &   & \checkmark  & Sept. 24th, 2017  & 6 &\hspace{-1em}  hours & 200 \\ \hline
  \end{tabular}
\end{sidewaystable}

\fi

\subsection{Overview of case studies}
We categorize case studies into four groups according to the type of sensing tasks involved and provide  an overview of each group in this section. We then describe how well the ParmoSense functions performed for each scenario. 

\vspace{0.5em}

\boldTitle{Urban-data collection during workshops (\UbiLectA--\OCUnivB)}
%
\UbiLectA--\OCUnivB are scenarios for workshop-style events, such as a Mapping-party~\cite{bib:haklay_OSM_2008, bib:hristova_lifeOfParty_ICWSM_2013, bib:mooney_mappingParty_IJSDIR_2015}, which is widely used in organizations such as OpenStreetMap.
The overview of each scenario is as follows:
\dlist{
  \item[\dltxt Scenario \UbiLectA--\Suita (Mapping-party)]\dlbr{0mm}
        These scenarios were aimed at collecting unmapped geographical data. In this event, we collected information on trees (names and positions) in our university campus.
  \item[\dltxt Scenario \UbiLectB (FixMyStreet)]\dlbr{0mm}
        This scenario was for collecting dynamic geographical data such as road breakages, graffiti, and street lamp failures by imitating the mechanism on FixMyStreet~\cite{bib:fixmystreet_org}.
  \item[\dltxt Scenario \Ikoma--\OCUnivB (Urban-planning)]\dlbr{0mm}
       These scenarios involved surveying existing POIs such as local buildings, public facilities, and nature, for urban planning.
}

Through the deployment of these scenarios, we obtained the following knowledge:
\ilist{
  \item Because \participants were willing to attend the event by themselves, we confirmed that the set of \MotivatingFuncs on ParmoSense are enough for getting sufficient contributions from the general public.
  \item The following \PSfunctions of \ProcessingFuncs worked effectively:
            \ilist{
              \item Data labeling function
              \item Cleansing function for unnecessary data
              \item Automatic mosaic function by face recognition
              \item Data export function
            }
}

\vspace{0.5em}

\boldTitle{Urban-data collection during sightseeing (\UbiLabA--\AIPB)}
%
\UbiLabA--\NICTB are scenarios for sightseeing PMS, e.g., with an information sharing tool for tourists, or an urban sensing system mimicking a tourist guide.
The overview of each scenario is as follows:
\dlist{
  \item[\dltxt Scenario \UbiLabA (Experience-sharing)]\dlbr{0mm}
    This scenario involved sharing discovered sightseeing spots among a group of tourists.
    To encourage positive posting among participants, we used points as the motivating function (F4).
  \item[\dltxt Scenario \NICTA (Tourist-guidance)]\dlbr{0mm}
    This scenario involved collecting data such as photos and behavior logs from tourists, while providing sightseeing information through a virtual tour guide~\cite{bib:hidaka_curation_smartcities2019}.
    We provided a data editing function (F6) for organizers to edit collected data after the event.
  \item[\dltxt Scenario \NICTB--\AIPB (Multi-type-requests)]\dlbr{0mm}
    The scenario was for comprehensively collecting environmental data of sightseeing spots, by requesting it in various ways from tourists.
    To encourage active contribution of participants, we provided all the available \MotivatingFuncs such as static/dynamic request (F3), points and coupons (F4). \figref{fig:kyoto-experiment} shows Screenshots of ParmoSense Client which are providing motivating functions including the dynamic points controls based on the demand for sensing tasks.
    \figref{fig:kyoto-human-behavior} visualizes participants' trajectories when they use ParmoSense which has/does not have \MotivatingFuncs (F4). The area marked with dashed lines shows motivating functions (weighted points) effectively work to change user’s behaviors by dynamically weighting points which can be earned. It suggests \MotivatingFuncs in ParmoSense could contribute to solving the spatio-temporal coverage issue of general PMS systems. The detail analysis of these case studies (\NICTB and \AIPA) are provided in our other papers~\cite{bib:shltn_gamification_smartcities2020, bib:shltn_tourism_mobiquitous2020}.
}

\fig{width=\columnwidth}{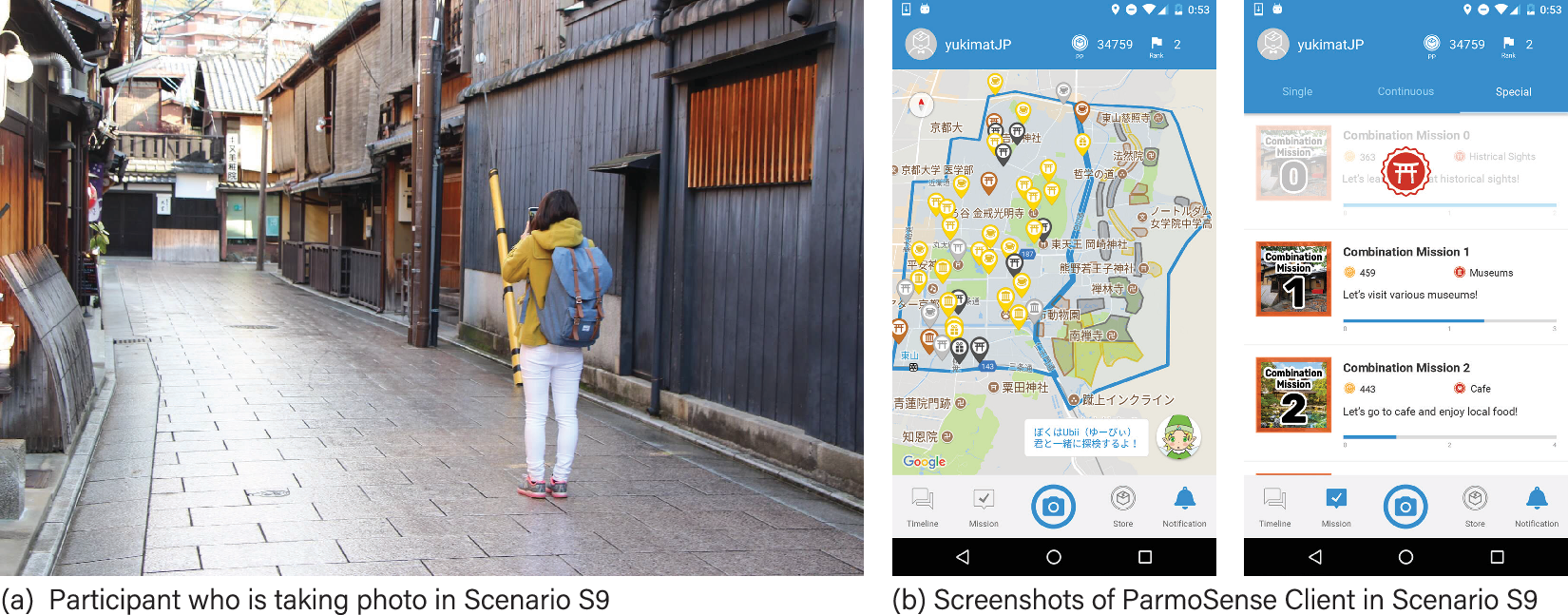}{\textbf{A photo of participant and screenshots of ParmoSense Client in Scenario \NICTB.}}{fig:kyoto-experiment}{\vspace{2mm}}

\fig{width=\columnwidth}{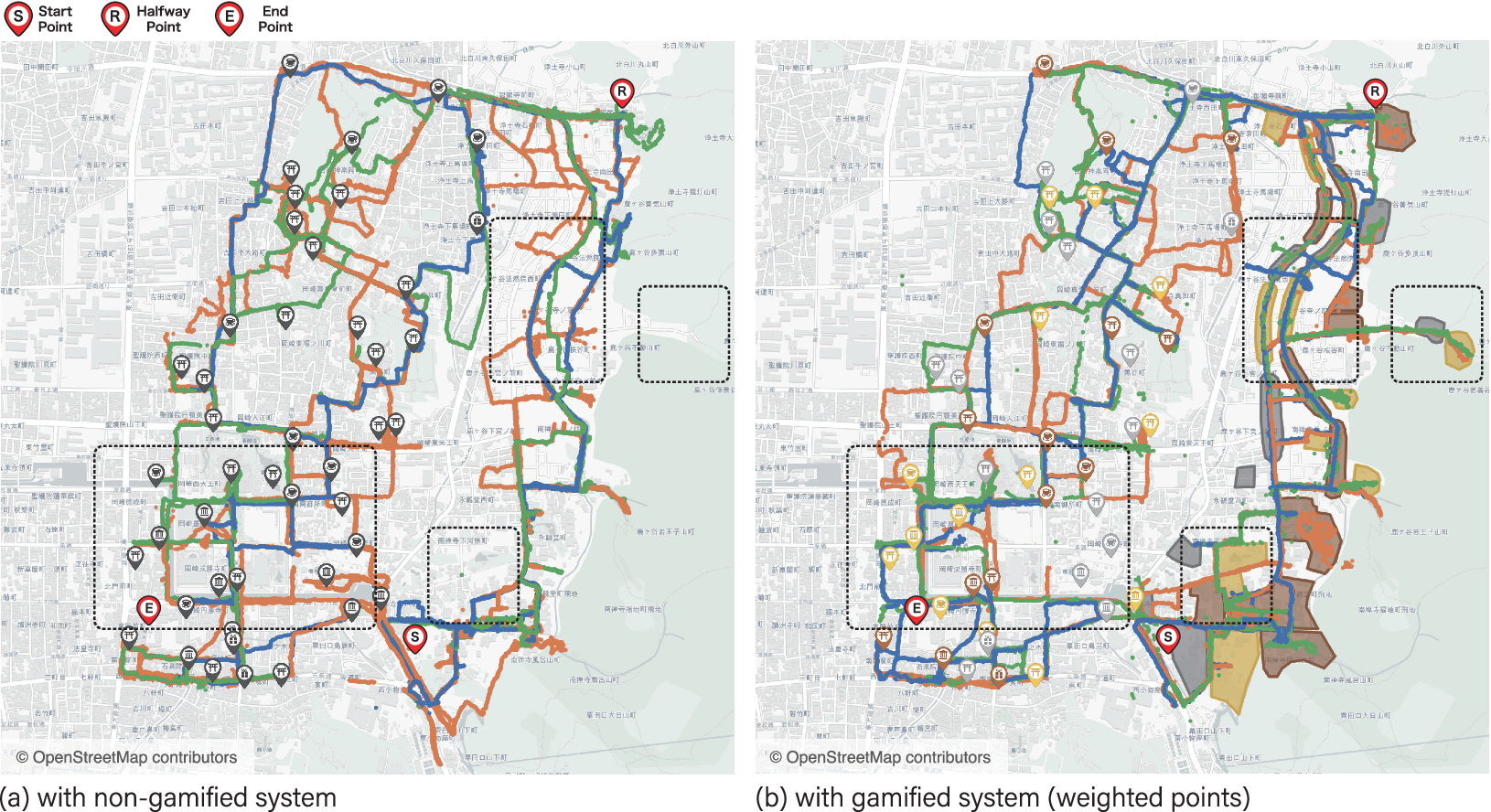}{\textbf{User's trajectories with/without \MotivatingFuncs in Scenario \NICTB.}}{fig:kyoto-human-behavior}{\vspace{2mm}}

Through these case studies, we observed the following on the effectiveness of the implemented \PSfunctions for scenarios belonging to this category:
\ilist{
  \item In the case of sightseeing, \MotivatingFuncs were essential because tourists participated in the scenario opportunistically.
  \item Collecting factors, e.g., points and sharing information with each other, 
         were more effective than the competitive factors, e.g., score and ranking, at motivating \participant's continuous participation because ``sightseeing'' was their primary purpose and collecting factors were more relevant.
}

\vspace{0.5em}

\boldTitle{Urban-data collection in daily life (\MThesis--\UbiLectC)}
%
The data suitable for collecting by PMS are often ``continuous'' and ``long-term'' data, because PMS is a sustainable sensing mechanism to realize comprehensive spatio-temporal data collection without any infrastructure due to the use of the many mobile devices dispersed in the city.
\MThesis and \UbiLectC were scenarios for collecting such long-term and continuous data.
The overview of each scenario is as follows:
\dlist{
  \item[\dltxt Scenario \MThesis (Static-request)]\dlbr{0mm}
    This scenario was for periodically collecting information that changes day by day, e.g., bulletin board information at a facility, and temperature of a place.
       To encourage participation, we set a limit on the number of participant contributions resulting from static requests (F3) that would be accepted, which is similar to the Audition mechanism~\cite{bib:ra_medusa_mobisys_2012}.
  \item[\dltxt Scenario \UbiLectC (Dynamic-request)]\dlbr{0mm}
    This scenario was for conducting questionnaires linked to location information by interrupting participants.
    Push notifications were used, and participants' location and behavior were taken into consideration, to provide dynamic requests (F3). No maps, rewards, or visualizations were included.
}

Through the case studies involving these scenarios, we concluded the following regarding the effectiveness of ParmoSense \PSfunctions intended for urban-data collection in daily life:
\ilist{
   \item In the case of \MThesis, due to the use of static requests through placing pins on a map, 
      the achievement of sensing depended on the active and continuous contribution of \participants themselves.
       The following \PSfunctions worked efficiently when motivating contribution:
        \ilist{
            \item Virtual rewards (e.g., point, ranking, level)
            \item Monetary incentives
            \item Visualization of contributions of the participant and of other \participants
        }
    \item Restricting the number of contributions on static requests was effective for encouraging \textit{continuous} participation because it prevented point inflation, but it also caused the leaving of some \participants.
    \item In the case of \UbiLectC, we found that contribution of \participants improved 
          when the timing of requests was based on participant's location and behavior, e.g., participants were more likely to respond to requests if the location of the sensing task was near to their present location.
}

\fig{width=0.8\columnwidth}{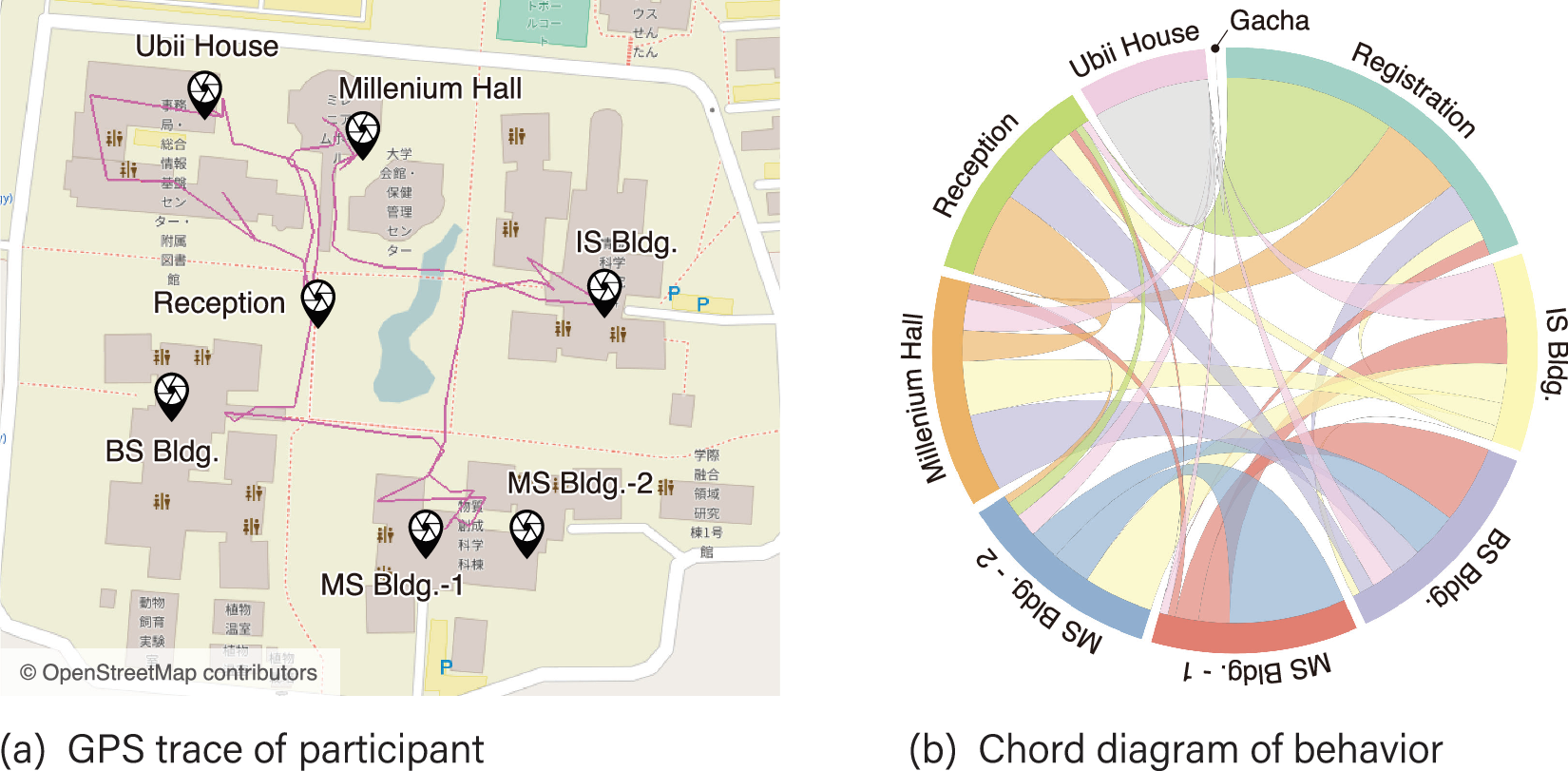}{\textbf{Result of human-behavior investigation in Scenario \CICP.}}{fig:human-behavior}{}

\vspace{0.5em}

\boldTitle{Human-behavior investigation on events (\AyaseA--\OsakaUnivB)}
%
\ParmoSense can not only collect urban environmental data, but also data of ``people'' existing in the city.
\AyaseA--\OsakaUnivB were scenarios created for investigating human behavior during various situations: events, daily life, sightseeing and so on.
The overview of each scenario is as follows:
\dlist{
  \item[\dltxt Scenario \AyaseA--\AyaseB (Stamp-rally)]\dlbr{0mm}
    These scenarios was for investigating the behavior of people who participate in the electronic ``stamp rally'' which is a process where visitors are instructed to go around to certain places, events etc. location~\cite{bib:rallyapp_jp}.
    A physical (electronic) stamp is put in a predetermined place, and when a visitor arrives, a stamp is stuck on a sheet (application).
    Reward is given based on the number of stamps accumulated. The rewards include coupons, prizes, points and ranking as non-monetary/monetary rewards (F4).
    We provided a data browsing function for visualizing \participants' behavior (F7).
   \figref{fig:human-behavior} shows examples of visualization of \participants' behavior.
  \item[\dltxt Scenario \UbiLabB (Sightseeing)]\dlbr{0mm}
     This scenario was for collecting \participants' movement data linked with the location information, to investigate the \participants' behavior tendencies during sightseeing.
     To continuously sense tourists' behavior, we used an implicit sensing function that ran in the background (F1).
  \item[\dltxt Scenario \OsakaUnivA--\OsakaUnivB (w/external-sensor)]\dlbr{0mm}
     These scenarios was used to collect \participants' movement and heart rate data linked to location information, to be used to construct a database for human behavior analysis. We added a function to connect a wearable sensor to external sensors (F1) via BLE, for collecting data.
}

Through these case studies, we learned the following about PMS for human-behavior investigation:
\ilist{
  \item To analyze the behavior of a person, there is a need to visualize behavior logs in various ways. We found that the functions provided by ParmoSense 
        such as GPS trace, and Chord diagram listings worked effectively.
  \item 
      Since these scenarios were aimed at sensing of human behavior affected by the specific function, it was necessary to suppress interference of elements, which is except for the function to be evaluated, to people.
      We confirmed that ParmoSense can minimize the number of unnecessary functions because of module-based design.
  \item  It is sometimes necessary to collect data with high frequency over a long term (\OsakaUnivA and \OsakaUnivB were scenarios that required acquiring the 9-axis sensor data at 100 Hz over several hours). In this case, we found that functions such as continuous data acquisition in the background worked efficiently.
}

\newcommand{\Q}[1]{{\texttt{\textbf{Q#1}}}\xspace}

\newcommand{\QueParticipantA}{\Q{1}}
\newcommand{\QueParticipantAwhy}{\Q{2}}
\newcommand{\QueParticipantB}{\Q{3}}
\newcommand{\QueParticipantBwhy}{\Q{4}}
\newcommand{\QueParticipantC}{\Q{5}}

\newcommand{\QueOrganizerA}{\Q{6}}
\newcommand{\QueOrganizerB}{\Q{7}}
\newcommand{\QueOrganizerC}{\Q{8}}
\newcommand{\QueOrganizerD}{\Q{9}}
\newcommand{\QueOrganizerE}{\Q{10}}
\newcommand{\QueOrganizerF}{\Q{11}}
\newcommand{\QueOrganizerG}{\Q{12}}

\newcommand{\nmTwo}{\mbox{}\\[-1.5em]}
\newcommand{\nmThree}{\mbox{}\\[-2em]}
\newcommand{\mrTwo}[1]{\multirow{2}{2em}{\nmTwo\centering #1}}
\newcommand{\mrThree}[1]{\multirow{3}{2em}{\nmThree\centering #1}}
\newcommand{\QFont}{\fontsize{9.5pt}{0pt}\selectfont}

\subsection{Survey and discussion}

In the following sections, we summarize and discuss the results of the subjective surveys held in each case study. We asked each \participant and \organizer questions about the usability and performance of \ParmoSense.

\subsubsection{Overview}

Participants in nine scenarios (\Suita, \Ikoma--\OCUnivB, \NICTB, \UbiLectC--\AyaseB) were given the questionnaire. These scenarios had many ordinary citizens participating, and include at least one motivating function. The total number of participants who answered the questionnaire was 201. About 70\% of participants were in their 20s, however, various age groups ranging 10s to 80s were targeted. 
Attributes of \participants are listed below:

\ilist{
  \item Citizens living in the local area, men and women of all ages (\Suita, \Ikoma, \CICP--\AyaseB)
  \item Undergraduate students majoring in architectural engineering in their 20s (unfamiliar with the information science) (\OCUnivA, \OCUnivB)
  \item Graduate students majoring in information science (\NICTB, \UbiLectC)
}

We also distributed questionnaires to the \organizers of five scenarios (\Suita, \Ikoma--\OCUnivB, \NICTB).
The skills of the organizers varied from those with high IT skills like IT engineer to those with low IT skills like students majoring in fields other than IT.
The attributes of the \organizers are listed below:
\ilist{
  \item Employee of regional public facility (\Suita)
  \item IT engineer (\Ikoma)
  \item Undergraduate students majoring in architectural engineering (\OCUnivA, \OCUnivB)
  \item Graduate students majoring in information science (\NICTB)
}

\newcommand{\QandA}[1]{\textit{``#1''}}
\newcommand{\ans}[1]{``#1''}

\subsubsection{Survey of participants}

To evaluate the usability of \ParmoSenseClient, we asked participants the questions which are listed in \tabref{tab:question_participants}.
we asked \QueParticipantA (\QandA{Was \ParmoSenseClient easy to use?}) with a 5-point Likert scale (5: very easy to use -- 1: very hard to use), and then asked for the reason in \QueParticipantAwhy.
The total number of answers was 196.
The breakdown of answers is shown in \tabref{tab:question_participants}.
The average score was 3.4 and about 40\% of \participants answered \QandA{easy to use.}

Participants who answered \QandA{easy to use} had the following to say:
\ilist{
  \item \QandA{We could operate intuitively because of simple application design.} (\Ikoma, \OCUnivB, \AyaseB)
  \item \QandA{Map visualization function is helpful for understanding the collected data having position information.} (\Ikoma, \OCUnivA, \OCUnivB)
}

Those who answered \QandA{hard to use} gave the following comments:
\ilist{
  \item \QandA{It was difficult to use \ParmoSense application because I was not used to smartphones.} (\Ikoma, \CICP)
  \item \QandA{\ParmoSense application was not suitable for long-term use because this application consumes battery more than expected.} (\Ikoma, \OCUnivA, \OCUnivB)
  \item \QandA{Unused functions should be invisible.} (\UbiLectC)
}

Overall, we found that using ParmoSense was easy for \participants who use smartphones on a daily basis regardless of age. In the case of local events where elderly persons also attended, some seniors felt that it was difficult to use a smartphone. However, this is not a particular problem of ParmoSense.
As the use case diversifies, the information displayed on the screen also diversifies.
Thus, to solve this issue without programming skills, it is necessary to modularize the screen configuration of the application and also to enable screen layout design in  \ScenarioTools according to the use case.

Next,  we asked \QueParticipantB (\QandA{Was the event using \ParmoSense fun?}) with a 5-point Likert scale (5: very fun -- 1: not fun at all) in target scenarios except for \UbiLectC.
Additionally, we also asked \QueParticipantBwhy to solicit free responses from participants on how motivating functions such as Reward and Visualization affected their sensing behavior.
The total number of answers was 103.
The breakdown of answers is shown in \tabref{tab:question_participants}.
The average score from the responses was 4.2 and about 80\% of \participants answered \QandA{fun.}

In \QueParticipantBwhy, we obtained the following comments:
\ilist{
   \item \QandA{It is pleasant to see the behavior of other people and other groups by checking the pins on the map and timeline in real time.} (\Ikoma, \MThesis)
  \item \QandA{Ranking function and leveling function makes it fun and encouraging.} (\Ikoma, \OCUnivA, \NICTB)
  \item \QandA{This application increased the fun of the stamp rally.} (\CICP, \AyaseB)\\[-2mm]
}

\begin{table}[t]
  \centering
  \caption{\textbf{Questionnaire items for participants and non-participants (\Suita, \Ikoma--\OCUnivB, \NICTB, \UbiLectC--\AyaseB).}}
  \label{tab:question_participants}
  \begin{tabular}{c|l|ccccc|c}\hline\hline
  \multirow{3}{2em}{\nmTwo\centering Item\\No.} 
       & \multirow{3}{17em}{\nmTwo\centering Questionnaire Detail}
       & \multicolumn{6}{c}{Answer}              \\ \cline{3-8}
      &       & \hw{2.2}{1}   & \hw{2.2}{2}   & \hw{2.2}{3}   & \hw{2.2}{4}   & \hw{2.2}{5} & \multirow{2}{3.5em}{\nmTwo\centering Average}  \\[-0.8em]
      &       & \fontsize{6.2pt}{0pt}\selectfont (disagree)  &         &         &         & \fontsize{6.2pt}{0pt}\selectfont (agree) &   \\ \hline
  \mrTwo{\QueParticipantA}
                        & \multirow{2}{17em}{\QFont\nmTwo Was \ParmoSenseClient easy to use?\\
                              \scriptsize (\Suita, \Ikoma--\OCUnivB, \NICTB, \UbiLectC--\AyaseB) }  
                         & 8            & 19             & 73               & 35            & 22      &  \multirow{2}{3.5em}{\nmTwo\centering 3.4}    \\[-0.4em]
                &       & \footnotesize (5.0\%)  & \footnotesize (11.9\%)    & \footnotesize (45.6\%)      & \footnotesize (21.9\%)  & \footnotesize (15.6\%)  &   \\ \hline
  \mrTwo{\QueParticipantAwhy}
                        & \multirow{2}{17em}{\QFont\nmTwo Why did you think so about the answer to \QueParticipantA? 
                              \scriptsize (\Suita, \Ikoma--\OCUnivB, \NICTB, \UbiLectC--\AyaseB) }
                        & \multicolumn{5}{c|}{\multirow{2}{12em}{\QFont\nmTwo\centering (Open-ended question)}} & \mrTwo{-}  \\[-0.4em]
                        &&&&&&&\\ \hline
  \mrTwo{\QueParticipantB}
                        & \multirow{2}{17em}{\QFont\nmTwo Was the event using ParmoSense fun?\\
                              \scriptsize (except \UbiLectC) }  
                        & -            &  4            & 14               & 42            &  43       &  \multirow{2}{3.5em}{\nmTwo\centering 4.2}      \\[-0.4em]
                &       & \footnotesize (0.0\%)   & \footnotesize (3.9\%)   & \footnotesize (13.6\%)      & \footnotesize (40.8\%)  & \footnotesize (41.7\%)     &\\ \hline
  \mrTwo{\QueParticipantBwhy}
                        & \multirow{2}{17em}{\QFont\nmTwo Why did you think so about the answer to \QueParticipantB? 
                              \scriptsize (except \UbiLectC) }
                        & \multicolumn{5}{c|}{\multirow{2}{12em}{\QFont\nmTwo\centering (Open-ended question)}} & \mrTwo{-}  \\[-0.4em]
                        &&&&&&&\\ \hline
  \mrThree{\QueParticipantC}
                        & \multirow{3}{17em}{\QFont\nmTwo What factors will encourage you to participate in experiments?
                              \scriptsize (Non-participants of \UbiLectC) }
                        & \multicolumn{5}{c|}{\multirow{3}{12em}{\QFont\nmTwo\centering (Open-ended question)}} & \mrThree{-}  \\[-0.4em]
                        &&&&&&& \\
                        &&&&&&& \\ \hline
  \end{tabular}
\end{table}

\subsubsection{Survey of non-participants}

We also provided questionnaires to non-participants in the free participation scenario (\UbiLectC) in order  to explore the reasons why they did not participate in the sensing tasks.

In scenario \UbiLectC, 35 of 118 candidates did not respond to the request to attend our experiment. 
We asked these 35 non-participants the open-ended question \QueParticipantC (\QandA{What factors will encourage you to participate in experiments?}).
The factors that promoted participation were related to ease of participation (17 people), battery consumption concerns (10 people), rewards (13 people), and benefit or convenience (12 people) respectively.
Furthermore, four people were concerned about privacy, such as \QandA{requiring personal information (e.g., e-mail address)} and \QandA{cannot be anonymous.}

Almost half of the non-participants pointed out that the number of steps they needed to take in order to participate was inconvenient. To minimize the steps to participate in multiple scenarios, we first required \participants to install \ParmoSenseClient. After installing the application, \participants read a QR code of each scenario in order to participate in that particular scenario. However, people who participated in only one scenario felt that this is complicated and that having a single, pre-configured application for a specific scenario might be easier.

Also, to make the registration and login process easier, we adopted Google Authentication (using Gmail address) and supported auto-login. This is a standard method used in many applications. However, some participants felt that our application could collect private data such as the e-mail address. In the future, we aim to address this by introducing an anonymous participation mode that does not require participant registration.

\renewcommand{\nmTwo}{\mbox{}\\[-1.8em]}
\renewcommand{\nmThree}{\mbox{}\\[-2.35em]}
\newcommand{\mrOne}[1]{\multirow{1}{2em}{\mbox{}\\[-1em]\centering #1}}

\begin{table}[t]
  \centering
  \caption{\textbf{Questionnaire items for organizers (\Suita, \Ikoma--\OCUnivB, \NICTB).}}
  \label{tab:question_organizers}
  \begin{tabular}{c|p{21.3em}|cccc|c}\hline\hline
      \multirow{3}{2em}{\nmTwo\centering Item\\No.}        & 
        \multirow{3}{21em}{\nmTwo\centering Questionnaire Detail}   & 
        \multicolumn{5}{c}{Answer}                \\ \cline{3-7}
          &       & \hw{2.8}{1}   & \hw{2.8}{2}   & \hw{2.8}{3}   & \hw{2.8}{4}  & \multirow{2}{3.5em}{\nmTwo\centering Average} \\[-0.8em]
          &       & \scriptsize (disagree)  &         &        & \scriptsize (agree)  &  \\ \hline
      \mrOne{\QueOrganizerA}
            & \QFont Was \ParmoSense easy to introduce to the event? 
            & -    & -    & 3      & 2     & 3.4\\ \hline
      \mrOne{\QueOrganizerB}
            & \QFont Could you collect the desired data?
            & - & - & 1 & 4  &  3.8\\\hline
      \mrTwo{\QueOrganizerC}
            & \QFont Were participants' motivation for attending to events
              improved by using ParmoSense? \small (e.g., ranking, coupons)
            & \mrTwo{-} & \mrTwo{-} & \mrTwo{3} & \mrTwo{2}  & \mrTwo{3.4} \\ \hline
      \mrOne{\QueOrganizerD}
          & \QFont Was \DataTools easy to use?   
            & -    & 1    & 2    & 2    & 3.2 \\ \hline
      \mrTwo{\QueOrganizerE}
            & \QFont Were the data outputted by the data output function
               easy to use secondary diversion and data processing?
             & \mrTwo{-}   & \mrTwo{1}    &  \mrTwo{3}      &  \mrTwo{1}   & \mrTwo{3.0}  \\ \hline
      \mrTwo{\QueOrganizerF}
            & \QFont Do you want to use \ParmoSense again in future similar events?
            & \mrTwo{-} & \mrTwo{-} & \mrTwo{2} & \mrTwo{3} & \mrTwo{3.6} \\ \hline
      \mrOne{\QueOrganizerG}
            & \QFont Why did you think so about the answer to \QueOrganizerF?   
             & \multicolumn{4}{c|}{\QFont(Open-ended question)} & -     \\ \hline
  \end{tabular}
  \vspace{-0em}
\end{table}

\subsubsection{Survey of organizers}

To evaluate the effect of \ParmoSense from the view of organizers and the usability of \ProcessingFuncs, we conducted a questionnaire with organizers.
The questions for organizers are listed in \tabref{tab:question_organizers}. \QueOrganizerA--\QueOrganizerF were answered with 4-point Likert scales (4: strongly agree -- 1: strongly disagree), \QueOrganizerG was a free-response question.

To evaluate the \organizer's burden indicated in \figref{fig:design_concept} -- \DistributingPhase, we asked about \QueOrganizerA (Ease of creating and distributing scenarios) to \organizers.
Two people answered \ans{4} and the other three answered \ans{3,} giving an average score of 3.4.
Although the IT skills of the \organizers were quite different, all of them felt that ParmoSense was easy to use for PMS.

Next, we asked \QueOrganizerB 
(Whether they were able to collect desired data)
and \QueOrganizerC 
(Performance of motivating function), 
to get feedback from \organizers about \figref{fig:design_concept} -- \SensingPhase.
Four organizers answered \ans{4} and one answered \ans{3} in \QueOrganizerB, 
while two answered \ans{4} and the other three answered \ans{3} in \QueOrganizerC.
Consequently, we can say that \ParmoSense allows \organizers to collect desired data and to motivate participants to contribute the sensing continuously, from a \organizer's objective perspective.

To evaluate \figref{fig:design_concept} -- \ProcessingPhase, 
we asked the organizers about the \QueOrganizerD (Usability of \DataTools) and \QueOrganizerE (Availability of output data).
Two organizers answered \ans{4} 
and two answered \ans{3} 
and another organizer answered \ans{2} 
for \QueOrganizerD, 
while one organizer answered \ans{4} 
and three answered \ans{3} 
and other one answered \ans{2} 
for \QueOrganizerE.
The average scores were  3.2 and 3.0 for \QueOrganizerD and \QueOrganizerE, respectively.
According to these results, we can confirm that the \DataTools of ParmoSense work substantially fine.

Additionally, we interviewed the organizers who answered \ans{2} in \QueOrganizerD and \QueOrganizerE, respectively.
The organizer whose answer for \QueOrganizerD is \ans{2} said \QandA{Unexpected data was downloaded when exporting data for each tag after tagging data.}
Therefore, as an additional question, we asked \QandA{Is it easy to use if data could be successfully downloaded?,} and we got the answer \QandA{I think so.}
This shows that although it is necessary to improve the flexibility in the export function, it seems that the usability of Web editor meets specific service standards.
Next, the organizer whose answer for \QueOrganizerE is \ans{2} said \QandA{When doing Web visualization, the operation became heavy and the PC froze many times.}
About 700 photos were collected in this organizer's event. This data amount was too large to display all photos at once. 
In the future, it is necessary to set an upper limit on the number of photos that can be displayed, or reduce the image size according to the number of images before uploading.

Finally, three organizers answered \ans{4}, two answered \ans{3} in \QueOrganizerF (Whether to use ParmoSense again).
When we asked the reason in \QueOrganizerG, the following answers were obtained:
\ilist{
  \item \QandA{It can find places where the participants are interested in.}
  \item \QandA{It can easily collect data with location information and can edit detailed information later. 
    In addition, by visualizing the data, participants can understand how the data is used intuitively.}
}

From these results, we found that ParmoSense is useful for human-behavior analysis and feedback of results to participants and it is easy to edit data.

%
%
%

\section{Conclusion}
\label{sec:conclusion}

Demands for the assessment of the urban environments through PMS systems are spread not only in the information science field, but also in a wide range of areas, such as urban planning and public services. In such areas, organizers are not always familiar with information technology. Therefore, \ParmoSense was designed to serve as a useful tool to collect the urban environmental information easily and flexible regardless of \organizers' IT skills.
\ParmoSense allows organizers to construct, distribute, and introduce various types of PMS system for various purposes by modularizing \PSfunctions and describing combinations and settings as ``scenario.''
In PMS systems that involve ordinary citizens for collecting data, motivating \participants is also important for continuous system operation.
Thus, \ParmoSense supported several \MotivatingFuncs as an incorporated feature, which was not supported on conventional platforms.

Through performance comparison in the perspective of varieties of functions and preparation workload with existing PMS platforms, we confirmed ParmoSense shows the best cost-performance. In addition, through 19 case studies over four years, we confirmed that ParmoSense could be compatible flexibly with various sensing tasks, motivate methods, and data processing.
Moreover, ParmoSense can suppress the burden of organizers and participants in system operation, and we found that it has higher cost performance than any conventional platforms.
On the other hand, we also found \ParmoSense has not been a ``magic bullet'' solution that can be used in every situation.
\ParmoSense should improve privacy and security by allowing \organizers to control them based on the sensing purpose.
In addition, to operate scenarios that involve a more significant number of \participants, robustness and scalability should be guaranteed based on distributed processing and prediction of score update and so on.

%
\section{Acknowledgments}
This research was funded by JST PRESTO (JPMJPR2039) and JSPS KAKENHI (JP19K24345 and JP19H01139).

%
\bibliographystyle{unsrt}
\bibliography{references}

\begin{thebibliography}{10}

\bibitem{bib:burke_participatory_2006}
J.~A Burke, D.~Estrin, M.~Hansen, A.~Parker, N.~Ramanathan, S.~Reddy, and M.~B
  Srivastava.
\newblock Participatory sensing.
\newblock {\em Center for Embedded Network Sensing}, 2006.

\bibitem{bib:campbell_acm_2006}
Andrew~T Campbell, Shane~B Eisenman, Nicholas~D Lane, Emiliano Miluzzo, and
  Ronald~A Peterson.
\newblock People-centric urban sensing.
\newblock In {\em Proceedings of the 2nd annual international workshop on
  Wireless internet}, page~18. ACM, 2006.

\bibitem{bib:paulos_citizenScience_2008}
Eric Paulos, R~Honicky, and Ben Hooker.
\newblock Citizen science: Enabling participatory urbanism.
\newblock {\em Urban Informatics: Community Integration and Implementation},
  2008.

\bibitem{bib:morishita_ubicomp_2015}
Shigeya Morishita, Shogo Maenaka, Nagata Daichi, Morihiko Tamai, Keiichi
  Yasumoto, Toshinobu Fukukura, and Keita Sato.
\newblock Sakurasensor: Quasi-realtime cherry-lined roads detection through
  participatory video sensing by cars.
\newblock {\em Proc. UBICOMP '15}, pages 695--705, 2015.

\bibitem{bib:konis_occupantMobileGateway_elsevier_2017}
K~Konis and M~Annavaram.
\newblock The occupant mobile gateway: a participatory sensing and
  machine-learning approach for occupant-aware energy management.
\newblock {\em Building and Environment}, 118:1--13, 2017.

\bibitem{bib:rolt_collega_iscc_2016}
C.~R.~De Rolt, R.~Montanari, M.~L. Brocardo, L.~Foschini, and J.~da~Silva~Dias.
\newblock Collega middleware for the management of participatory mobile health
  communities.
\newblock In {\em 2016 IEEE Symposium on Computers and Communication (ISCC)},
  pages 999--1005, June 2016.

\bibitem{bib:heggen_ubicomp_2012}
Scott Heggen.
\newblock Integrating participatory sensing and informal science education.
\newblock In {\em Proceedings of the 2012 ACM Conference on Ubiquitous
  Computing}, UbiComp '12, pages 552--555. ACM, 2012.

\bibitem{bib:kok_journalofPerCom_2014}
Kok-Leong Ong, Simone Leao, and Adam Krezel.
\newblock Participatory sensing and education: Helping the community mitigate
  sleep disturbance from traffic noise.
\newblock {\em International Journal of Pervasive Computing and
  Communications}, 10(4):419--441, 2014.

\bibitem{bib:matsuda_ubicomp_2014}
Yuki Matsuda and Ismail Arai.
\newblock A safety assessment system for sidewalks at night utilizing
  smartphones' light sensors.
\newblock In {\em Proceedings of the 2014 ACM International Joint Conference on
  Pervasive and Ubiquitous Computing: Adjunct Publication}, pages 115--118.
  ACM, 2014.

\bibitem{bib:kanjo_NoiseSPY_2010}
Eiman Kanjo.
\newblock Noisespy: a real-time mobile phone platform for urban noise
  monitoring and mapping.
\newblock {\em Mobile Networks and Applications}, 15(4):562--574, 2010.

\bibitem{bib:maisonneuve_noise_2010}
Nicolas Maisonneuve, Matthias Stevens, and Bartek Ochab.
\newblock Participatory noise pollution monitoring using mobile phones.
\newblock {\em Information Polity}, 15(1):51, 2010.

\bibitem{bib:mendez_pSense_2011}
Diego Mendez, Alfredo~J Perez, Miguel Labrador, Juan~Jose Marron, et~al.
\newblock P-sense: A participatory sensing system for air pollution monitoring
  and control.
\newblock In {\em Pervasive Computing and Communications Workshops (PerCom
  Workshops), 2011 IEEE International Conference on}, pages 344--347. IEEE,
  2011.

\bibitem{bib:zheng_uAir_2013}
Yu~Zheng, Furui Liu, and Hsun-Ping Hsieh.
\newblock U-air: When urban air quality inference meets big data.
\newblock In {\em Proceedings of the 19th ACM SIGKDD international conference
  on Knowledge discovery and data mining}, pages 1436--1444. ACM, 2013.

\bibitem{bib:fixmystreet_org}
{FixMyStreet.org}.
\newblock {FixMyStreet Platform}.
\newblock \url{http://fixmystreet.org/}, 2018.

\bibitem{bib:ferreira_aware_frontiers_2015}
Denzil Ferreira, Vassilis Kostakos, and Anind~K Dey.
\newblock Aware: mobile context instrumentation framework.
\newblock {\em Frontiers in ICT}, 2(6):1--9, 2015.

\bibitem{bib:haoyi_sensus_ubicomp_2016}
Haoyi Xiong, Yu~Huang, Laura~E. Barnes, and Matthew~S. Gerber.
\newblock Sensus: A cross-platform, general-purpose system for mobile
  crowdsensing in human-subject studies.
\newblock In {\em Proceedings of the 2016 ACM International Joint Conference on
  Pervasive and Ubiquitous Computing}, UbiComp '16, pages 415--426. ACM, 2016.

\bibitem{bib:ra_medusa_mobisys_2012}
Moo-Ryong Ra, Bin Liu, Tom~F. La~Porta, and Ramesh Govindan.
\newblock Medusa: A programming framework for crowd-sensing applications.
\newblock In {\em Proceedings of the 10th International Conference on Mobile
  Systems, Applications, and Services}, MobiSys '12, pages 337--350. ACM, 2012.

\bibitem{bib:funf_social_fMRI_Percom_2011}
Nadav Aharony, Wei Pan, Cory Ip, Inas Khayal, and Alex Pentland.
\newblock {Social fMRI}: Investigating and shaping social mechanisms in the
  real world.
\newblock {\em Pervasive and Mobile Computing}, 7(6):643--659, 2011.

\bibitem{bib:sakamura_minaqn_ubicomp_2015}
Mina Sakamura, Tomotaka Ito, Hideyuki Tokuda, Takuro Yonezawa, and Jin
  Nakazawa.
\newblock Minaqn: Web-based participatory sensing platform for citizen-centric
  urban development.
\newblock In {\em Adjunct Proceedings of the 2015 ACM International Joint
  Conference on Pervasive and Ubiquitous Computing and Proceedings of the 2015
  ACM International Symposium on Wearable Computers}, UbiComp/ISWC'15 Adjunct,
  pages 1607--1614. ACM, 2015.

\bibitem{bib:mishima_kokopin_maed_2013}
M.~Mishima, T.~Matsumoto, S.~Takano, and O.~Matsuda.
\newblock Kokopin app: A mobile platform for biogeography.
\newblock In {\em Proceedings of the 2nd ACM International Workshop on
  Multimedia Analysis for Ecological Data}, MAED '13, pages 35--40. ACM, 2013.

\bibitem{bib:tangmunarunkit_ohmage_acmTran_2015}
H.~Tangmunarunkit, C.~K. Hsieh, B.~Longstaff, S.~Nolen, J.~Jenkins, C.~Ketcham,
  J.~Selsky, F.~Alquaddoomi, D.~George, J.~Kang, Z.~Khalapyan, J.~Ooms,
  N.~Ramanathan, and D.~Estrin.
\newblock Ohmage: A general and extensible end-to-end participatory sensing
  platform.
\newblock {\em ACM Transactions on Intelligent Systems and Technology},
  6(3):38:1--38:21, 2015.

\bibitem{bib:brunette_opendatakit_hotmobile_2013}
Waylon Brunette, Mitchell Sundt, Nicola Dell, Rohit Chaudhri, Nathan Breit, and
  Gaetano Borriello.
\newblock Open data kit 2.0: Expanding and refining information services for
  developing regions.
\newblock In {\em Proceedings of the 14th Workshop on Mobile Computing Systems
  and Applications}, HotMobile '13, pages 10:1--10:6. ACM, 2013.

\bibitem{bib:wang_GPSelector_www}
Jiangtao Wang, Yasha Wang, Leye Wang, and Yuanduo He.
\newblock Gp-selector: a generic participant selection framework for mobile
  crowdsourcing systems.
\newblock {\em World Wide Web}, 21:759--782, 2017.

\bibitem{bib:arakawa_gamification_2016}
Yutaka Arakawa and Yuki Matsuda.
\newblock Gamification mechanism for enhancing a participatory urban sensing:
  survey and practical results.
\newblock {\em Journal of Information Processing}, 24(1):31--38, 2016.

\bibitem{bib:gamification_co_fatigue_2012}
{Gamification Co}.
\newblock Gamification pitfalls: Badge fatigue and loyalty backlash.
\newblock
  \url{http://www.gamification.co/2012/09/12/gamification-pitfalls-badge-fatigue-and-loyalty-backlash/},
  2012.

\bibitem{bib:Francesco_incentive_2015}
Francesco Restuccia, Sajal~K Das, and Jamie Payton.
\newblock Incentive mechanisms for participatory sensing: Survey and research
  challenges.
\newblock {\em ACM Transactions on Sensor Networks}, 2015.

\bibitem{bib:luis_reverseAuction_IEEE_2012}
L.~G. Jaimes, I.~Vergara-Laurens, and M.~A. Labrador.
\newblock A location-based incentive mechanism for participatory sensing
  systems with budget constraints.
\newblock In {\em 2012 IEEE International Conference on Pervasive Computing and
  Communications}, pages 103--108, 2012.

\bibitem{bib:Gao_incentive_2015}
Hui Gao, Chi~Harold Liu, Wendong Wang, Jianxin Zhao, Zheng Song, Xin Su, Jon
  Crowcroft, and Kin~K Leung.
\newblock {A Survey of Incentive Mechanisms for Participatory Sensing}.
\newblock {\em IEEE Communications Surveys Tutorials}, 17(2):918--943, 2015.

\bibitem{bib:ogie_incentiveMechanisms_springer_2016}
R.~I. Ogie.
\newblock Adopting incentive mechanisms for large-scale participation in mobile
  crowdsensing: from literature review to a conceptual framework.
\newblock {\em Human-centric Computing and Information Sciences}, 6(1):24,
  2016.

\bibitem{bib:ioannis_monetaryincentives_taylor2012}
Ioannis Krontiris and Andreas Albers.
\newblock Monetary incentives in participatory sensing using multi-attributive
  auctions.
\newblock {\em International Journal of Parallel, Emergent and Distributed
  Systems}, 27(4):317--336, 2012.

\bibitem{bib:zichermann_oreilly_2011}
Gabe Zichermann and Christopher Cunningham.
\newblock {\em Gamification by design: Implementing game mechanics in web and
  mobile apps}.
\newblock {O'Reilly Media, Inc.}, 2011.

\bibitem{bib:deterding_mindtrek_2011}
Sebastian Deterding, Dan Dixon, Rilla Khaled, and Lennart Nacke.
\newblock From game design elements to gamefulness: defining gamification.
\newblock In {\em Proceedings of the 15th International Academic MindTrek
  Conference: Envisioning Future Media Environments}, pages 9--15. ACM, 2011.

\bibitem{bib:groh_ulm_2012}
Fabian Groh.
\newblock Gamification: State of the art definition and utilization.
\newblock {\em Institute of Media Informatics Ulm University}, 39, 2012.

\bibitem{bib:wang_mobilecrowdsourcing_willey2016}
Yufeng Wang, Xueyu Jia, Qun Jin, and Jianhua Ma.
\newblock Mobile crowdsourcing: framework, challenges, and solutions.
\newblock {\em Concurrency and Computation: Practice and Experience},
  29(3):e3789, 2016.
\newblock e3789 CPE-15-0470.R1.

\bibitem{bib:ueyama_gamification_2014}
Yoshitaka Ueyama, Morihiko Tamai, Yutaka Arakawa, and Keiichi Yasumoto.
\newblock Gamification-based incentive mechanism for participatory sensing.
\newblock In {\em Pervasive Computing and Communications Workshops (PerCom
  Workshops), 2014 IEEE International Conference on}, pages 98--103, 2014.

\bibitem{bib:funf_org}
{MIT Media Lab}.
\newblock Funf - open sensing framework.
\newblock \url{http://funf.org/}, 2017.

\bibitem{bib:mqtt_org}
{OASIS}.
\newblock {MQTT.org}.
\newblock \url{http://mqtt.org/}, 2014.

\bibitem{bib:openstreetmap_org}
{OpenStreetMap Foundation}.
\newblock {OpenStreetMap}.
\newblock \url{http://www.openstreetmap.org/}, 2018.

\bibitem{bib:google_earth}
{Google LLC}.
\newblock {Google Earth}.
\newblock \url{https://www.google.com/earth/}, 2018.

\bibitem{bib:cesiumjs_org}
Cesium.
\newblock {CesiumJS}.
\newblock \url{https://cesium.com/cesiumjs/}, 2018.

\bibitem{bib:hidaka_curation_smartcities2019}
Masato Hidaka, Yuki Kanaya, Shogo Kawanaka, Yuki Matsuda, Yugo Nakamura,
  Hirohiko Suwa, Manato Fujimoto, Yutaka Arakawa, and Keiichi Yasumoto.
\newblock On-site trip planning support system based on dynamic information on
  tourism spots.
\newblock {\em Smart Cities}, 3(2):212--231, 2020.

\bibitem{bib:shltn_gamification_smartcities2020}
Shogo Kawanaka, Yuki Matsuda, Hirohiko Suwa, Manato Fujimoto, Yutaka Arakawa,
  and Keiichi Yasumoto.
\newblock Gamified participatory sensing in tourism: An experimental study of
  the effects on tourist behavior and satisfaction.
\newblock {\em Smart Cities}, 3(3):736--757, 2020.

\bibitem{bib:shltn_tourism_mobiquitous2020}
Shogo Kawanaka, Juliana Miehle, Yuki Matsuda, Hirohiko Suwa, Keiichi Yasumoto,
  and Minker Wolfgang.
\newblock Design and evaluation on task allocation interfaces in gamified
  participatory sensing for tourism.
\newblock In {\em Proceedings of the Workshop on Artificial Intelligence for
  Mobile and Ubiquitous Communication System}, MobiQuitous '20, page 1–6.
  ACM, 2020.

\bibitem{bib:haklay_OSM_2008}
M.~Haklay and P.~Weber.
\newblock Openstreetmap: User-generated street maps.
\newblock {\em IEEE Pervasive Computing}, 7(4):12--18, Oct 2008.

\bibitem{bib:hristova_lifeOfParty_ICWSM_2013}
Desislava Hristova, Giovanni Quattrone, Afra Mashhadi, and Licia Capra.
\newblock The life of the party: Impact of social mapping in openstreetmap.
\newblock In {\em International AAAI Conference on Web and Social Media}, pages
  234--243, 2013.

\bibitem{bib:mooney_mappingParty_IJSDIR_2015}
Peter Mooney, Marco Minghini, and Frances Stanley{-}Jones.
\newblock Observations on an openstreetmap mapping party organised as a social
  event during an open source {GIS} conference.
\newblock {\em {IJSDIR}}, 10, 2015.

\bibitem{bib:rallyapp_jp}
{FaithCreates Inc.}
\newblock {Rally}.
\newblock \url{https://rallyapp.jp/en/}, 2017.

\end{thebibliography}

\end{document}